\documentclass[aps,pre,twocolumn,groupedaddress]{revtex4-2}

\usepackage{graphicx}
\usepackage{caption}
\usepackage{dcolumn}
\usepackage{bm}
\usepackage{caption}
\usepackage{subcaption}
\usepackage{siunitx}
\usepackage{hyperref}
\usepackage{natbib}
\usepackage{amsmath}
\usepackage{xcolor}
\usepackage{tikz}
\usetikzlibrary{shapes}

\usepackage{xcolor} 

\usepackage{hyperref}

\newcommand{\greentrngle}{\raisebox{0.5pt}{\tikz{\node[regular polygon,scale=0.3,regular polygon, regular polygon sides=3,fill=green!100!green,rotate=0](){};\draw[-,green,solid,line width = 1pt](-2.0mm,0.0mm) --(-0.5mm,0.0mm);\draw[-,green,solid,line width = 1pt](0.5mm,0.0mm) --(2mm,0.0mm)}}}

\newcommand{\magentatrngle}{\raisebox{0.5pt}{\tikz{\node[regular polygon,scale=0.3,regular polygon, regular polygon sides=3,fill=magenta!100!magenta,rotate=0](){};\draw[-,magenta,solid,line width = 1pt](-2.0mm,0.0mm) --(-0.5mm,0.0mm);\draw[-,magenta,solid,line width = 1pt](0.5mm,0.0mm) --(2mm,0.0mm)}}}

\newcommand{\redcirc}{\raisebox{0pt}{\protect\tikz{\filldraw[-,red,solid,line width = 1pt](2.375mm,0.6mm)circle(2.2pt);\draw[-,red,solid,line width = 1pt](0mm,0.6mm) --(1.5mm,0.6mm);\draw[-,red,solid,line width = 1pt](3.3mm,0.6mm) --(5mm,0.6mm)}}}

\newcommand{\cyandiamond}{\raisebox{0pt}{\protect\tikz{\node[diamond, cyan, draw, scale=0.4, fill=cyan](){}; \draw[-,cyan,solid,line width = 1pt](-2.0mm,0.0mm) --(-0.5mm,0.0mm);\draw[-,cyan,solid,line width = 1pt](0.5mm,0.0mm) --(2mm,0.0mm)}}}

\newcommand{\invtriblue}{\raisebox{0pt}{\tikz{\node[draw,scale=0.3,regular polygon, regular polygon sides=3,fill=blue!45!blue,rotate=180](){}; \draw[-,blue,solid,line width = 1pt](-2.0mm,0.0mm) --(-0.5mm,0.0mm);\draw[-,blue,solid,line width = 1pt](0.5mm,0.0mm) --(2mm,0.0mm) }}}

\newcommand{\blacksquare}{\raisebox{0.5pt}{\tikz{\node[draw,scale=0.4,regular polygon, regular polygon sides=4,fill=black](){}; \draw[-,black,solid,line width = 1pt](-2.0mm,0.0mm) --(-0.5mm,0.0mm);\draw[-,black,solid,line width = 1pt](0.5mm,0.0mm) --(2mm,0.0mm)}}}

\newcommand{\textfrac}[2]{\ensuremath{#1/#2}}

\begin{document}


\title{\textbf{Energy non-equipartition in vibrofluidized particles} 
}%

\author{Alok Tiwari}%
\email{alok.tiwari@iitb.ac.in}
\author{Manaswita Bose}%
 \email{manaswita.bose@iitb.ac.in}
\affiliation{Department of Energy Science and Engineering, Indian Institute of Technology Bombay, Mumbai, India.
}%

\author{V. Kumaran}%
\affiliation{%
	Department of Chemical Engineering, Indian Institute of Science Bangalore, Bengaluru 560012, India
	}%

\date{\today}

\begin{abstract}
The aim of the present work is to investigate the influence of the realistic model parameters 
for particle
interactions, specifically the spring stiffness
coefficient for the tangential force between particles
on the energy equipartition in a vibrofluidized system. To achieve this, a three-dimensional vertically vibrated granular system consisting of spherical particles is simulated using the discrete element method (DEM) implemented in the open-source software LAMMPS. Interparticle and wall-particle interactions are determined using the linear-spring dashpot model. Simulations are performed for particles ranging from nearly perfectly smooth to nearly perfectly rough. Two different values for the ratio of the tangential to normal spring stiffness coefficient $\kappa$ ($2/7$ and $3/4$) are chosen for most of the simulations. The ratio of the translational to the rotational kinetic energy ($K$) monotonically decreases with an increase in the friction coefficient, $\mu$ for $\kappa=2/7$; however, for $\kappa = 3/4$, after an initial reduction with $\mu$, $K$ increases and plateaus at $\approx 5$, indicating the absence of equipartition of energy between the translational and rotational modes. Further simulations performed for $0.67 \le \kappa < 1$ confirm non-equipartition of energy for particles with a very high friction coefficient.

\end{abstract}

\maketitle


\section{\label{sec:Intro} Introduction}
The equipartition theorem states that the kinetic energy is equally distributed among all degrees of freedom in a fluid \citep{Reif}; however, the equipartition of energy deviates in gases in round vessels \citep{Naplekov2023}, bio-molecules \citep{Eastwood2010}, laser-cooled atoms \citep{Afek2020}, non-spheroidal molecules \citep{ErpenbeckandCohen1988}, granular mixtures \citep{Paolotti2003, Wildman2002, FeitosaandMenon2002, Puzyrev2024}, homogeneously cooling systems \citep{Trittel2024},  and granular gases with rough particles \citep{McNamara1998}. A seemingly simple system of a vibro-fluidized smooth particles deviates from equipartition of energy \citep{Kumaran1998pre} and anisotropy in the fluctuating kinetic energy $T_{x,y,z} = \frac{1}{2}\langle(u_{x,y,z}-\langle u\rangle_{x,y,z})^2\rangle$ is observed. The anisotropy in $T_{x,y,z}$ in a vertically vibrated granular system is due to the fact that the fluctuating kinetic energy is transferred from the bottom plate to the particles in the vertical direction. The energy is then distributed in the other two directions due to subsequent inter-particle interactions. The isotropic mean squared fluctuating kinetic energy of the particles $(T_o)$, which is obtained equating the rate of energy input to the system due to bottom-wall particle collision and rate of dissipation due to inelastic inter-particle collisions at the leading order in a moment expansion method \citep{Kumaran1998jfm}, scales as $\frac{U_o^2}{N\,d^2\left(1-e_n^2\right)}$, where $U_o$ is the wall velocity, $N$ is the number density, $d$ the particle diameter and $e_n$ the normal coefficient of restitution. The difference in the $T_{x}, T_y$ and $T_z$ is maximum near the bottom wall and monotonically decreases along the bed height and the $T_{x,y,z}$ asymptotically approaches $T_o$ for $N\,d^2\left(1-e_n^2\right) << 1$ \citep{Kumaran1998jfm, Sunthar1999}. The behaviour differs when dissipation due to air drag is considered. 

In an assembly of realistic granular particles, the partitioning of energy between the translational and rotational modes depends on the surface roughness (\cite{Rao2008} and references therein). In the limit of the nearly smooth particles, the rotational and translational kinetic energies are independently balanced, and the kinetic energy is not equally distributed between the rotational and the translational degrees of freedom. In the other limit of nearly perfectly rough particles, the partition of kinetic energy depends on the particle inelasticity and the surface roughness quantified in terms of the normal ($e_n$) and the rotational ($\beta$) coefficient of restitution \citep{Rao2008}. McNamara and Luding \citep{McNamara1998} defined a ratio $R = \frac{\Delta E^o}{\Delta \overline{E} + \Delta E^o}$ to quantify the partition of energy between the translational and rotational modes. Through event-driven simulations, they showed that $R$ weakly depends on the coefficient of restitution for $\beta \approx 0$. In the energy conserving limits, i.e., $\beta\sim -1 \,\,\mathrm{or} \,\,+1$, $R$ is independent of $e_n$. Grasselli et al \cite{Grasselli2015} studied the anisotropy in a 2-D vibro-fluidized granular bed in micro-gravity. They determined the rotational coefficient of restitution, the ratio of the rotational to the translational kinetic energy ($R_T$), and observed that the anisotropy in the fluctuating kinetic energy depends on the area fraction. Castillo et. al \cite{Castillo2020} discussed the departure from equipartition in the context of a granular system in a magnetically levitated bed.

Although the non-equipartition of kinetic energy in granular systems is widely observed, Nichol and Daniels \citep{Nichol2012} reported nearly equipartition of energy between the translational and rotational modes for a dense bimodal mixture subject to bidirectional periodic excitation on an air table. In their experiment, the granular assembly is excited by electromagnetic bumpers on three walls, with the fourth wall serving as a constraint.

Potiguar \citep{Potiguar2021} performed numerical simulations for the experimental set-up discussed in \citep{Nichol2012}. They used a linear spring dashpot model \citep{Cundall1979} to determine the contact force between the colliding disk-shaped particles, with the spring stiffness constant, $k_n=5\times 10^4 \frac{mg}{d}$, and $\gamma$, the dissipation coefficient, as two parameters. The tangential force is determined from the sliding friction coefficient ($\mu$). Simulations were performed for a wide range of $\gamma$, resulting in $0.2<e_n <0.9$ and for $\mu = 0.5$. They observed that the ratio of translational to rotational kinetic energy depends on the number density, coefficient of restitution, frequency, and magnitude of the energy injected.

The literature indicates that the non-equipartition of energy in a granular system depends on several factors, including particle properties, dissipation mechanisms, and particle number density.
The objective of the present work is to systematically investigate the effect of the friction coefficient and the ratio of the tangential to the normal stiffness coefficients on the partition of fluctuating kinetic energy between the translational and rotational modes of vibro-fluidized particles using the Discrete Element Method (DEM) and analyze the results in the framework proposed in \citep{McNamara1998, Rao2008}. To that end, simulations are performed using the open-source software LAMMPS with large values of spring stiffness constant, $k_n > 10^7 \frac{mg}{d}$ \citep{Reddy2010, Tiwari2024} to ensure binary collisions.

\section{\label{sec:method}Methodology}

\subsection{Background theory}
\label{sec:theory}

In the simplest hard-sphere model the collisions are characterized by two parameters: the normal  ($e_n = -\frac{v_n'}{v_n}$) and the rotational coefficients of restitution $(\beta = -\frac{\Vec{v}_s'\cdot\hat{k}_s}{\Vec{v}_s\cdot{k}_s})$, where, $v_n$ is the component of the relative velocity of particles along the line joining the centres of particles, and $\Vec{v}_{s} = \Vec{v}_{ij} -  \Vec{v}_{n} + \frac{d}{2}\left (\hat{r}_{ij} \times \Vec{\omega}_{ij} \right )$ is the slip velocity of the point of contact. The primed quantities represent the post-collision properties. The particle indices are $i$ and $j$, $\hat{r}_{ij}$ is the unit vector drawn from particle centre of $i$ to the centre of $j$, the subscript $n$ refers to the normal direction, $\Vec{v}_{ij}$ is the relative velocity of the particle $i$ with respect to $j$, and $\Vec{\omega}_{ij}$ is the sum of angular velocities about their respective centers, $\hat{k}_s$ is the unit vector in the slip velocity direction, defined as $\hat{k}_s = \frac{\vec{v}_s}{|\vec{v}_s|}$ \cite{Walton1993}.

The changes in the translational, rotational, and total kinetic energy, during a collision, are expressed by Equations ~\ref{eq:delta_te} -- \ref{eq:delta_total} \citep{Rao2008}:


\begin{align} \label{eq:delta_te}
    \Delta \left(\frac{1}{2}v'^2\right) = -\frac{1}{4}(1-e_n^2)(\hat{r}_{ij}\cdot \Vec{v}_{ij})^2 \nonumber 
    \\ - \eta_2 \left ( \hat{r}_{ij} \times \vec{G}\right ) \cdot \left ( \hat{r}_{ij} \times \vec{G}\right ) \nonumber  \\ + \eta_2^2 \left | \hat{r}_{ij} \times \vec{G} \right |^2 
- \eta_2 \left( \Vec{\omega}_{ij}\cdot \left ( \hat{r}_{ij} \times \vec{G} \right ) \right )
\end{align}


\begin{align}\label{eq:delta_re}
    \Delta \left(\hat{I}\omega'^2\right) = \eta_2 \left( \Vec{\omega}_{ij}\cdot \left ( \hat{r}_{ij} \times \vec{G} \right ) \right ) \nonumber 
    \\ + \frac{\eta_2^2}{\hat{I}} \left ( \hat{r}_{ij} \times \Vec{G}\right) \cdot \left ( \hat{r}_{ij} \times \Vec{G}\right )
\end{align}


\begin{align} \label{eq:delta_total}
    \Delta E=\Delta \left(\frac{1}{2}v'^2\right) + \Delta \left(\hat{I}\omega'^2\right) =  -\frac{1}{4}(1-e_n^2)(\hat{r}_{ij}\cdot \Vec{v}_{ij})^2  - \nonumber
    \\ \frac{\hat{I}}{1+\hat{I}}\frac{1-\beta^2}{4}\left ( \hat{r}_{ij} \times \Vec{G}\right ) \ \cdot \left ( \hat{r}_{ij} \times \Vec{G}\right )
\end{align}

\noindent where $\Delta \left(\frac{1}{2}v'^2\right)$ = $\frac{1}{2} (v_i^2 - v_i'^2) + \frac{1}{2}  (v_j^2 - v_j'^2)$, $\eta_2 = \frac{1}{2}(1+\beta)\hat{I}/(1+\hat{I})$, $\Vec{G} = \Vec{v}_{ij} + \vec{r}_{ij} \times (\vec{\omega}_i + \vec{\omega}_j)$, $\hat{I} = 4I/md^2$, and $I$ is the moment of inertia. Mass($m$) and diameter ($d$) of the particles are used as scaling parameters. The translational and rotational velocities are normalized with $U_o$ and $2U_o/d$, where $U_o = 2\pi A f$ is the maximum velocity of the vibrating base. The kinetic energy is conserved for a perfectly elastic ($e_n = 1$) collision between two perfectly smooth ($\beta = -1$) particles. For perfectly elastic particles with rough surfaces, the dissipation of energy is solely due to friction between particles and depends only on $\beta$. Otherwise, the change in the kinetic energy is a function of $e_n$ and $\beta$. The term $ E_{\text{exch}}
 = \eta_2 \left| \Vec{\omega}_{ij}\cdot \left ( \hat{r}_{ij} \times \vec{G} \right ) \right |$ accounts for the gain in the rotational energy compensating for the loss in translational energy and is a function of $\beta$. The ratio of the transfer of energy from the translation to the rotational mode to the energy dissipation $\Theta = \frac{ E_{\text{exch}}}{\Delta E}$, in general, depends on both $e_n$ and $\beta$. For perfectly elastic particles, $\Theta$ is an explicit function of the rotational coefficient of restitution ($\beta$). $\beta$ depends on the friction coefficient ($\mu$), the impact angle ($\gamma$), and the normal coefficient of restitution ($e_n$). The collision is said to be sliding if $-1\le \beta \le 0$. in this regime, $\beta = -1 + \frac{7}{2} \mu (1+e_n)\cot{\gamma}$ \citep{Walton1986}. In the stick-slip regime when $0 < \beta \le 1$, $\beta$ is a complex function of the material properties \citep{Kosinski2020}. Luding and McNamara\cite{McNamara1998} showed the dependence of the distribution of mean fluctuating kinetic energy in the rotational and the translational model on $\beta$ using event-driven simulations. They have shown that the distribution is independent of the normal coefficient of restitution.

\subsection{Simulation Method}

\begin{figure}[h]
\centering
\includegraphics[scale=0.35]{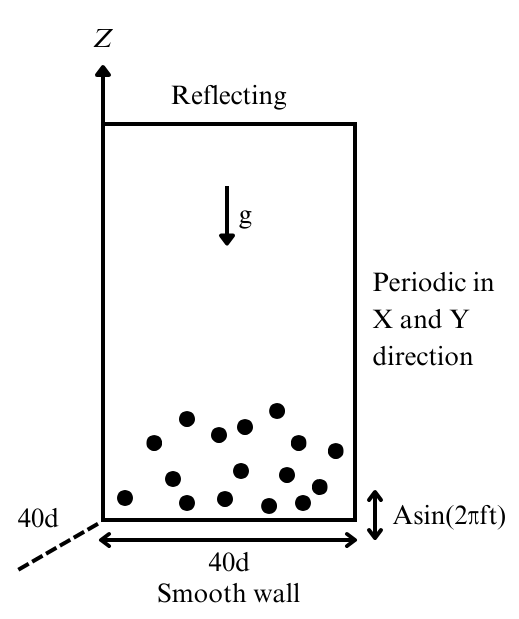}
\caption{Schematic of the simulation domain.} \label{fig:sim_domain_c5}
\end{figure}

Figure \ref{fig:sim_domain_c5} shows the schematic representation of the computational domain. The domain is periodic in the gravity normal direction. The upper wall is placed at a height of ${1000d}$ to mimic a semi-infinite domain. The bottom wall vibrates sinusoidally with the maximum energy of ${U_{o}^2 = 4\pi^2 A^2 f^2}$, where $f$ and $A$ are the frequency and amplitude of the vibration, respectively. The base frequency is maintained constant at $f^* = 2.26\sqrt{d_p/g}$. 

The amplitude is varied between ${0.3d \leq A \leq 0.7d}$, resulting in the non-dimensional acceleration ($\Gamma = 4\pi^2Af^2/g$) in the range $ 60 \leq \Gamma \leq 140 $ \citep{Eshuis2007}. Simulations are performed for $0\le \mu \le 10$, and a wide range of $\epsilon=Nd^2(1-e_n^2)$, where $N$ is the number density of particles (number per unit base-area)

The linear spring dashpot (LSD) model is used to determine the contact force (Equations presented in Sec I of \citep{SupMat}). The normal spring stiffness constant ($k_n= 10^8 \frac{mg}{d}$) is selected to ensure that $\frac{t_c}{t_f} \ll 1$, where $t_c$ is the contact time and $t_f$ is the average time between two successive collisions \citep{Silbert2001, Reddy2010, Tiwari2024}. Simulations are performed for two different values of $\kappa=\frac{k_t}{k_n}$, where, $\kappa =$ $\frac{2}{7}, \frac{3}{4}$ \citep{Silbert2001, Tiwari2024}. The viscous dissipation coefficient $\gamma_n$ is set such that $e_n$ varies between 0.85 - 1 \citep{Schafer1996}. A wide range of friction coefficient $\mu$ ($0\le \mu \le 10$) is used in the present study. A very large value of $\mu$ is included in the simulation to mimic a nearly perfectly rough case \citep{Rao2008}.

Simulations are performed in two stages. First, the simulation is performed for $ 7 \times 10^7 t_c$, where $t_c = 3.1 \times 10^{-4}\sqrt{\frac{d}{g}}$, is the time spent at contact. 
The time step of $\Delta t = \frac{t_c}{10}$ is used at this stage. Once the fluidized particles' total kinetic energy reached a steady state, simulations were run with a smaller time step, $\Delta t = t_{c}/100$. Instantaneous linear and angular velocities of particles obtained from the DEM simulations are further analyzed to determine the profiles of the mean fluctuating translational $\left(T_{(x,y,z)} = \frac{1}{2}\langle(v_{(x,y,z)}-\langle v_{(x,y,z)}\rangle)^2 \rangle \right)$ and rotational $\left(T_{R,(x,y,z)} = \frac{\hat{I}}{2}\langle(\Omega_{(x,y,z)}-\langle \Omega_{(x,y,z)}\rangle)^2\right)$ kinetic energy. In a vibro-fluidized bed, there is no mean linear and angular velocity, and the simulation results confirm that both $\langle v_{(x,y,z)}\rangle \rangle$and  $\langle \Omega_{(x,y,z)}\rangle)$ are $\approx 0$. The linear and angular velocities are normalized as $\vec{v} = \vec{V}/{U_o}$ and $\vec{\Omega} = d\vec{\omega}/2U_o$. Here, $U_{o} = 2 \pi A f$ is the maximum velocity of the vibrating base. The particle's vertical position is measured relative to the vibrating base.

The bed height averaged mean translational and rotational fluctuating kinetic energy ($\mathrm{KE_T} =\frac{1}{2\,h}\int_{0}^{h}\left \langle \left \| \Vec{v}\right \|^2 \right \rangle dz $, $\mathrm{KE_R} =\frac{\hat{I}}{2\,h}\int_{0}^{h}\left \langle \left \| \Vec{\Omega}\right \|^2 \right \rangle dz$) are also determined. More than $10^3$ configurations are used to determine the ensemble averages represented within $\langle\,\rangle$.

\section{Results}\label{sec2}
\subsection{Profiles of the mean fluctuating kinetic energy}
\begin{figure*}
    \centering
    \subfloat[a][]{%
    \includegraphics[scale=0.25]{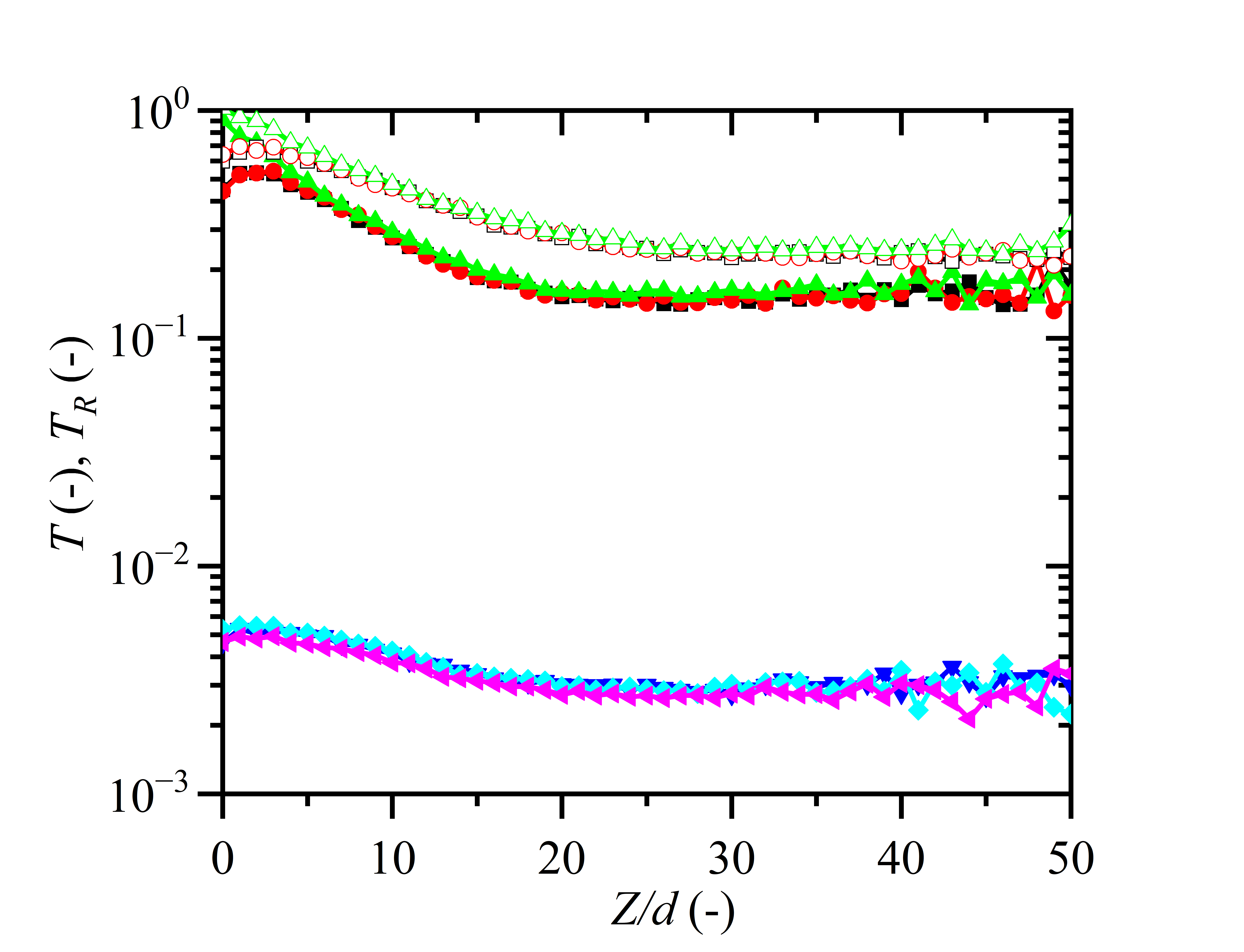}%
    \label{fig:mu0p01k2by7ep95}%
    }
    \subfloat[b][]{\includegraphics[scale=0.25]{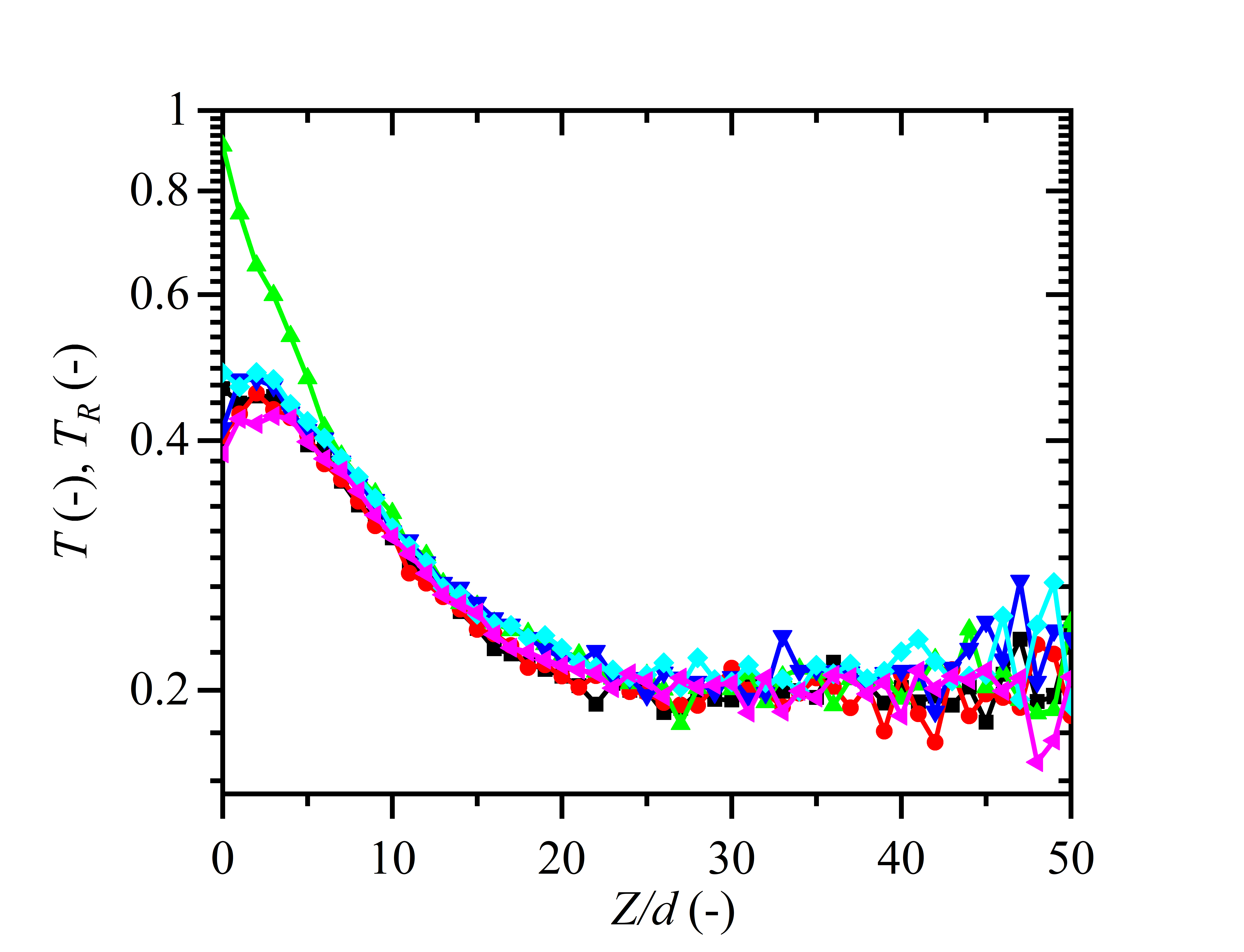}\label{fig:mu10k2by7ep95}}
    
    \subfloat[c][]{\includegraphics[scale=0.25]{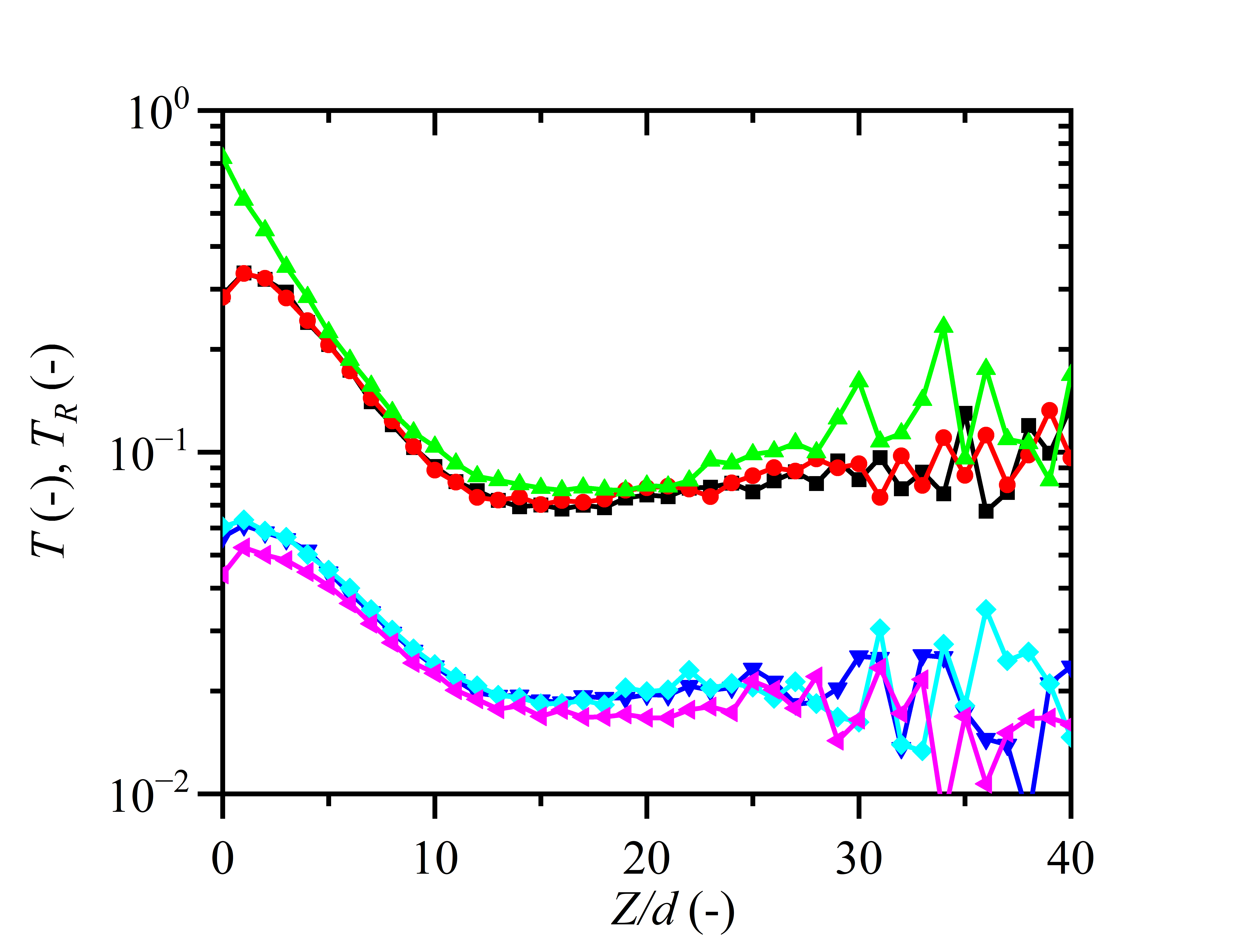}\label{fig:mu10k3by4ep95}}
    \subfloat[d][]{\includegraphics[scale=0.25]{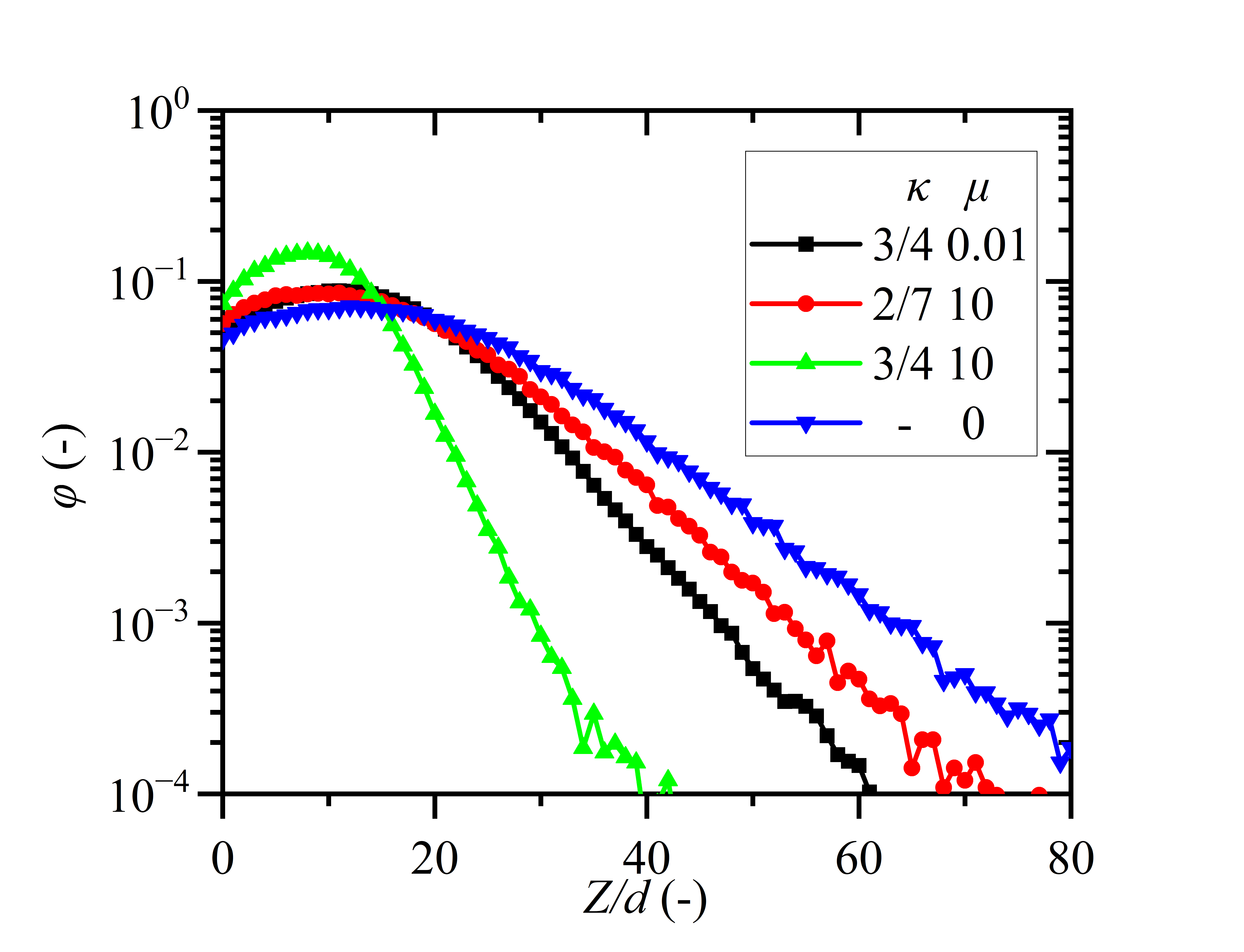}\label{fig:volfrac}}

    \caption{Profiles for the translational ($T_{(x,y,z)}$) and rotational ($T_{R,(x,y,z)}$) fluctuating kinetic energy for (a) $\mu = 0$ (hollow symbols) and $\mu = 0.01$ (filled symbols), with $\kappa = 2/7$, $e_n = 0.95$; (b) $\mu = 10$, $\kappa = 2/7$, $e_n = 0.95$; and (c) $\mu = 10$, $\kappa = 3/4$, $e_n = 0.95$.
Symbols: \protect\blacksquare, \protect\redcirc, and \protect\greentrngle\ denote the $T_x$, $T_y$, and $T_z$, respectively. \protect\invtriblue, \protect\cyandiamond, and \protect\magentatrngle\ represent $T_{R,x}$, $T_{R,y}$ and $T_{R,z}$, respectively. In (a), only translational granular temperature components are shown for $\mu = 0$ (hollow symbols), as rotational motion is not present in the frictionless case.
(d) Profile of the solid volume fraction corresponding to the specified values of $\kappa$ and $\mu$. The equilibrium temperature ($T_o$) obtained from the slope of the exponential fit of the volume fraction $-\frac{gz}{T_o}$ closely agrees with the equilibrium translation granular temperature, in those cases where equipartition was not observed.} 
    \label{fig:TnuProfiles}
\end{figure*}

Figures ~\ref{fig:mu0p01k2by7ep95} and ~\ref{fig:mu10k2by7ep95}   show the vertical profiles for $T_{(x,y,z)}$ and $T_{R,(x,y,z)}$ for $Nd^2(1-e_n^2)=0.39$, $e_n=0.95$, $\kappa=2/7$, $\mu = 0.01$ and a very high value of friction coefficient $(\mu = 10)$ respectively. 
Figure ~\ref{fig:mu0p01k2by7ep95} also plots the $T_{(x,y,z)}$ for perfectly smooth sphere ($\mu=0$). $T_z$ monotonically decreases with height and reaches the equilibrium temperature ($T_o$) beyond $\frac{z}{d}=30$.
The velocity fluctuations are anisotropic near the base, $T_z > T_x, T_y$, consistent with earlier
studies \citep{Kumaran1998jfm}. However, equipartition of translational fluctuating kinetic energy among the three degrees of freedom was achieved at $\frac{z}{d} \approx 10$ for the smooth particles. The temperature $T_{(x,y,z)}$ for nearly perfectly smooth particles ($\mu=0.01$) shows a very similar trend as shown by the perfectly smooth particles with $\mu = 0$. There is equipartition for the translational energy 
at $\frac{z}{d} \approx 25$. Profiles for $T_{R,(x,y,z)}$ show a monotonic decrease to an equilibrium value, with height. Also, $T_{R,(x,y,z)}$ is $\mathcal{O}(10^{-2})$ smaller than $T_{(x,y,z)}$ in this case.

For slightly rough particles ($ \mu = 0.01$, and $\beta \approx -1$), the translational and rotational energy conservation equations are nearly independently satisfied as $\eta_2\ll 1$ (Eqs ~\ref{eq:delta_te} and \ref{eq:delta_re}).

On the other hand, for nearly perfectly rough particles ($\beta \approx 1$), $\eta_2 \rightarrow\frac{\hat{I}}{1+\hat{I}}$, the maximum possible value, which leads to a strong coupling between the conservation of the translational and rotational fluctuating kinetic energy during a contact. Figure ~\ref{fig:mu10k2by7ep95} shows that $T_{(x,y,z)}$ and $T_{R,(x,y,z)}$ have equal values within numerical accuracy for $\frac{z}{d} > 10$.
 Analysis of the collisions with $\kappa=\frac{2}{7}$ for particles with $\mu > 0.5$ confirmed that the rotational coefficient of restitution, $\beta \approx +1$. Figure ~\ref{fig:mu10k3by4ep95} shows that there is a significant difference in the equilibrium values of $T$ and $T_R$, this is because for $\kappa=\frac{3}{4}$, $\beta < +1$ \citep{Tiwari2024}. The value of $\beta$ depends on the nature of the particle contact defined in \citep{Maw1976}. The fraction of the particle-contacts following the gross sliding determined using the method outlined in \citep{Tiwari2024}, for different $\kappa$ is presented in \citep{SupMat}. 
 Figure~\ref {fig:volfrac} shows the solid volume fraction in log-linear scale. For all cases, the profiles of the solid volume fraction show a maximum followed by an exponential decay with a slope $\frac{-gz}{T_o}$, where $T_o$ is the equilibrium temperature. The volume fraction profile of particles with very high surface roughness $\mu = 10$ with $\kappa = \frac{3}{4}$ shows a steeper decay than the other three cases, which is consistent with the results in Figure ~\ref{fig:mu10k3by4ep95}.

\begin{figure*}
    \centering
    \subfloat[a][]{\includegraphics[scale=0.25]{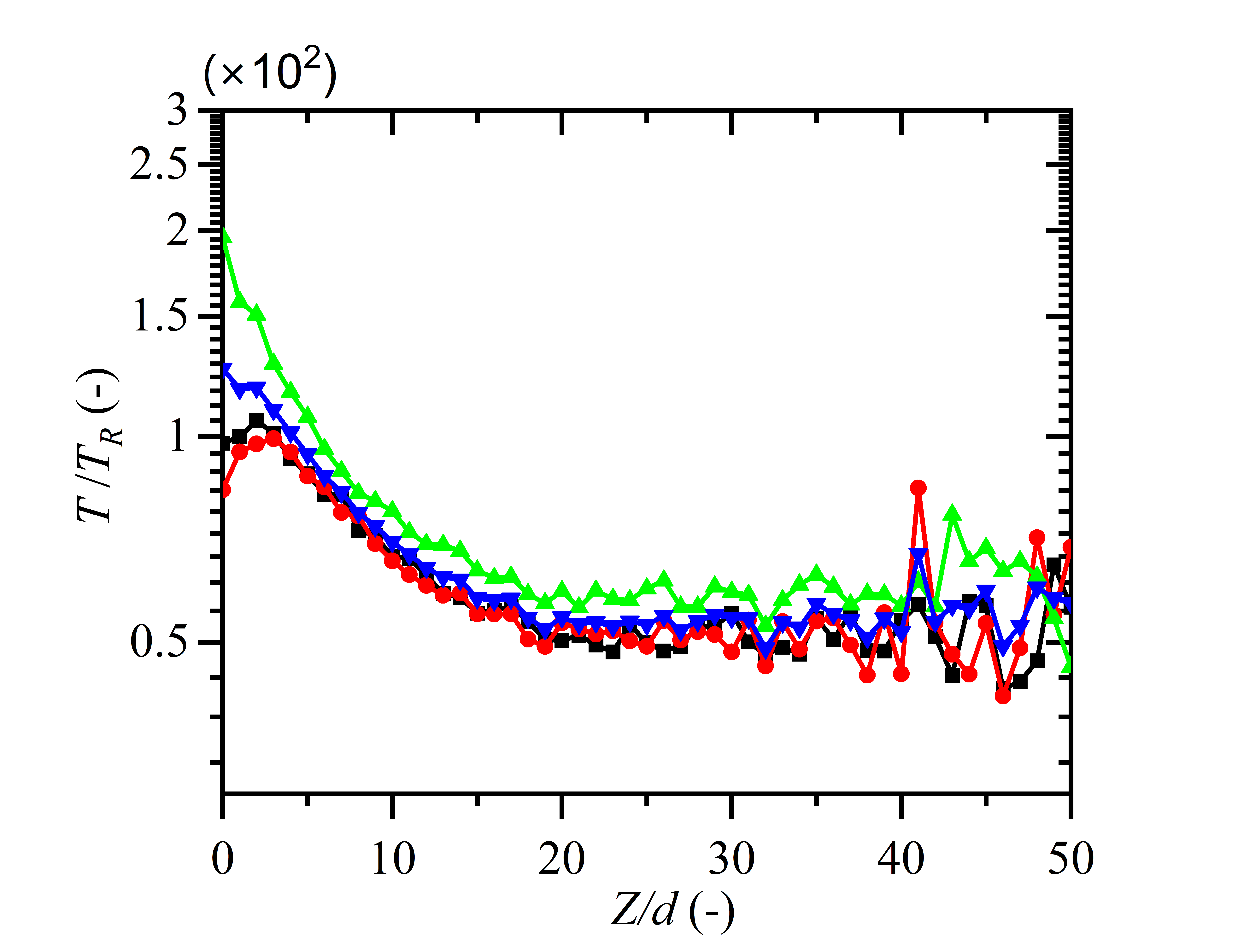}\label{fig:mu_0.01_ratios}}
    \subfloat[b][]{\includegraphics[scale=0.25]{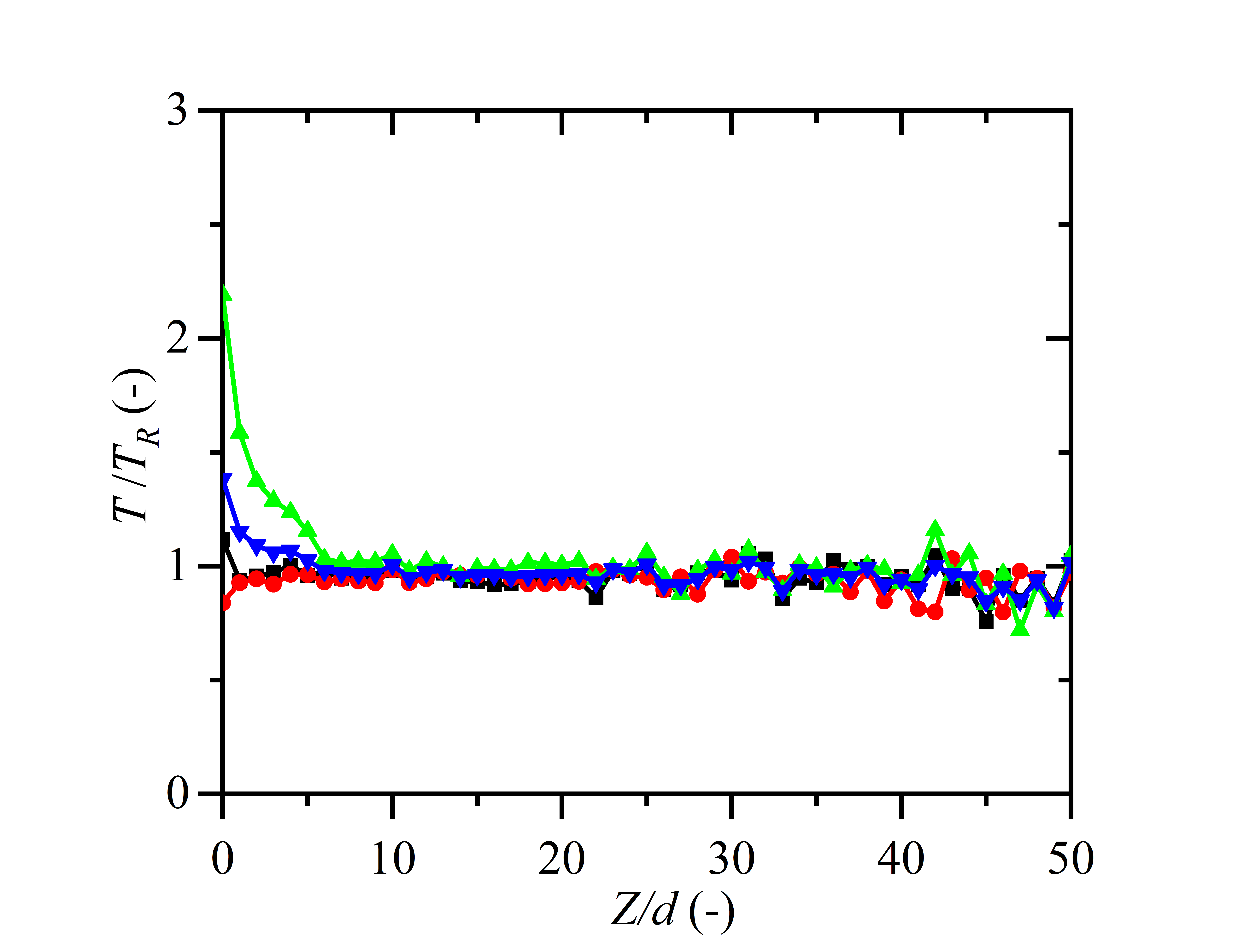}\label{fig:mu_10_2_7_ratios}}

    \subfloat[c][]{\includegraphics[scale=0.25]{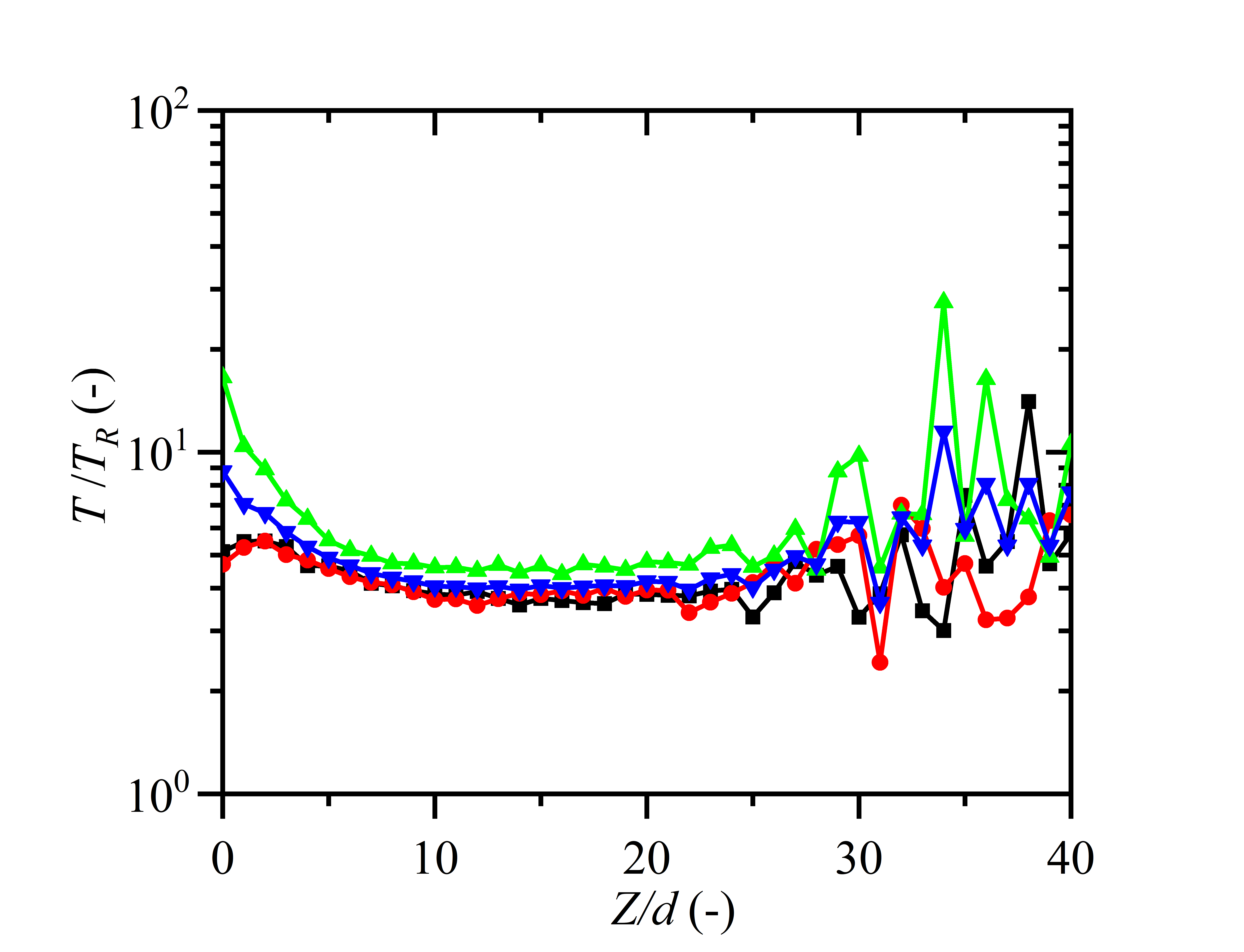}\label{fig:mu_10_3_4_ratios}}
 
    \caption{Profiles for the ratio of Translational $T_{(x,y,z)}$ and rotational $T_{R,(x,y,z)}$ fluctuating kinetic energy for (a) $\mu = 0.01$ with $\kappa = 2/7$, $e_n = 0.95$; (b) $\mu = 10$, $\kappa = 2/7$, $e_n = 0.95$; and (c) $\mu = 10$, $\kappa = 3/4$, $e_n = 0.95$. \protect\blacksquare: $\textfrac{T_x}{T_{R,x}}$, \protect\redcirc: $\textfrac{T_y}{T_{R,y}}$, \protect\greentrngle: $\textfrac{T_z}{T_{R,z}}$, \protect\invtriblue: $\textfrac{T}{T_R}$} 
    \label{fig:temp_ratios}
\end{figure*}

Figure \ref{fig:temp_ratios} plot $\frac{T_{(x,y,z)}}{T_{R,(x,y,z)}}$ and $\frac{T(z)}{T_R(z)}$, where $T_{(R)}(z)  = \frac{1}{3}\left(T_{x(R)}(z)+T_{y(R)}(z)+T_{z(R)}(z)\right)$. For $\kappa = \frac{2}{7}$, $\mu = 0.01$, the ratio of the $T$ and $T_R$ at equilibrium $\frac{T_o}{T_{Ro}} = \mathcal{O}(10^2)$ and for $\mu=10$ the ratio is close to unity. Figure \ref{fig:mu_10_3_4_ratios} shows that $\frac{T_o}{T_{Ro}}$ plateaus at $\approx 5$ for $\kappa = \frac{3}{4}$ and $\mu = 10$.

 The equilibrium ``granular temperature'' $(T_o)$ is obtained by equating the rate of energy input due to the collision with the vibrating wall and the rate of dissipation due to inelastic and frictional contacts truncated at the leading order. For perfectly smooth and nearly perfectly rough particles, $T_o$ can be expressed by Eqs \ref{eq:Tosmooth} and \ref{eq:Torough} respectively \citep{Kumaran1998jfm, Rao2008}:
 \begin{eqnarray}
 \label{eq:Tosmooth}
 T_o &=&\frac{\sqrt{2}U_o^2}{\pi Nd^2(1-e_n^2)}\\
 \label{eq:Torough}
 T_o &=&\frac{\sqrt{2}U_o^2}{\pi Nd^2[(1-e_n^2)+(1-\beta^2)]}
 \end{eqnarray}
 For nearly perfectly smooth particles in a vibro-fluidized bed where there is no mean angular velocity ($\langle \Omega\rangle = 0$), $T_o$ and $T_{R,o}$ are distinct, the expression for the $T_o$ depends on $T_{R,o}$:
 \begin{equation}
 \label{eq:TTrRough}
     T_o =\frac{\sqrt{2}U_o^2}{\pi Nd^2[(1-e_n^2)+\frac{\hat{I}}{1+\hat{I}}(1-\beta^2) + \frac{T_{R,o}}{T_o}\frac{1}{1+\hat{I}}(1-\beta^2)]}
 \end{equation}

 \begin{figure*}
     \centering
     \subfloat[a][]
     {\includegraphics[scale=0.25]{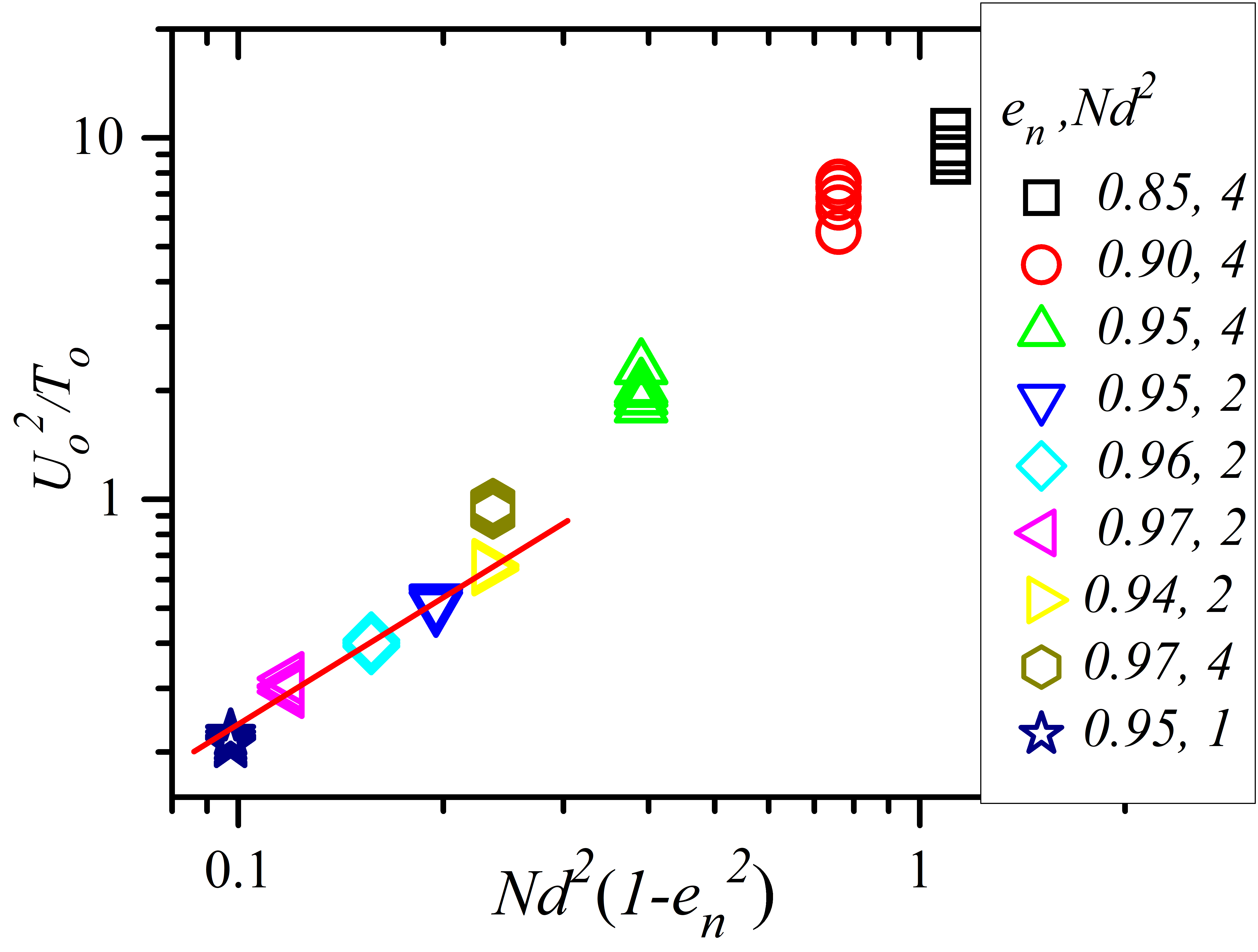}\label{fig:smoothscale}}
     \subfloat[b][]
     {\includegraphics[scale=0.25]{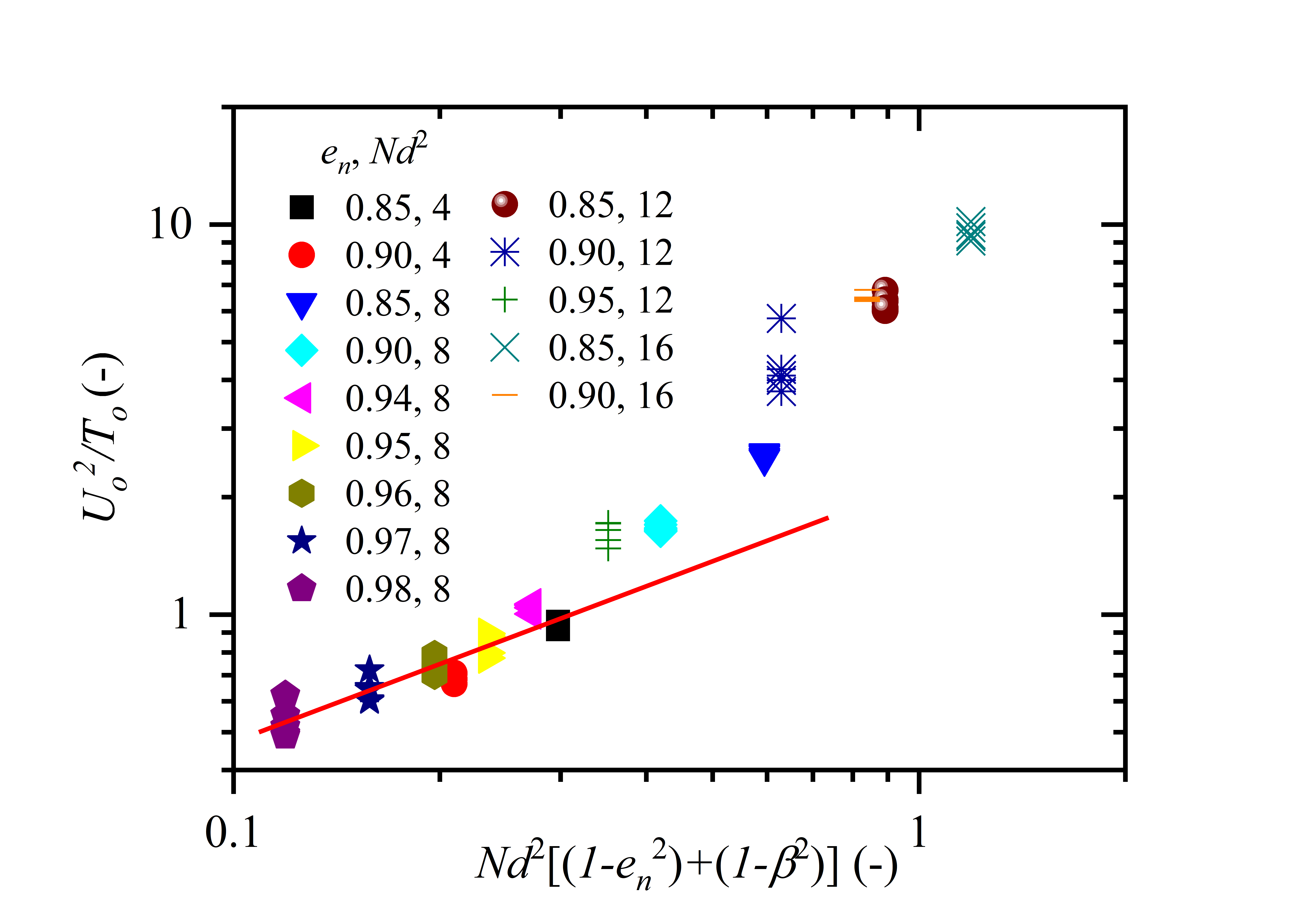}\label{fig:roughscale}}

     \subfloat[c][]
     {\includegraphics[scale=0.25]{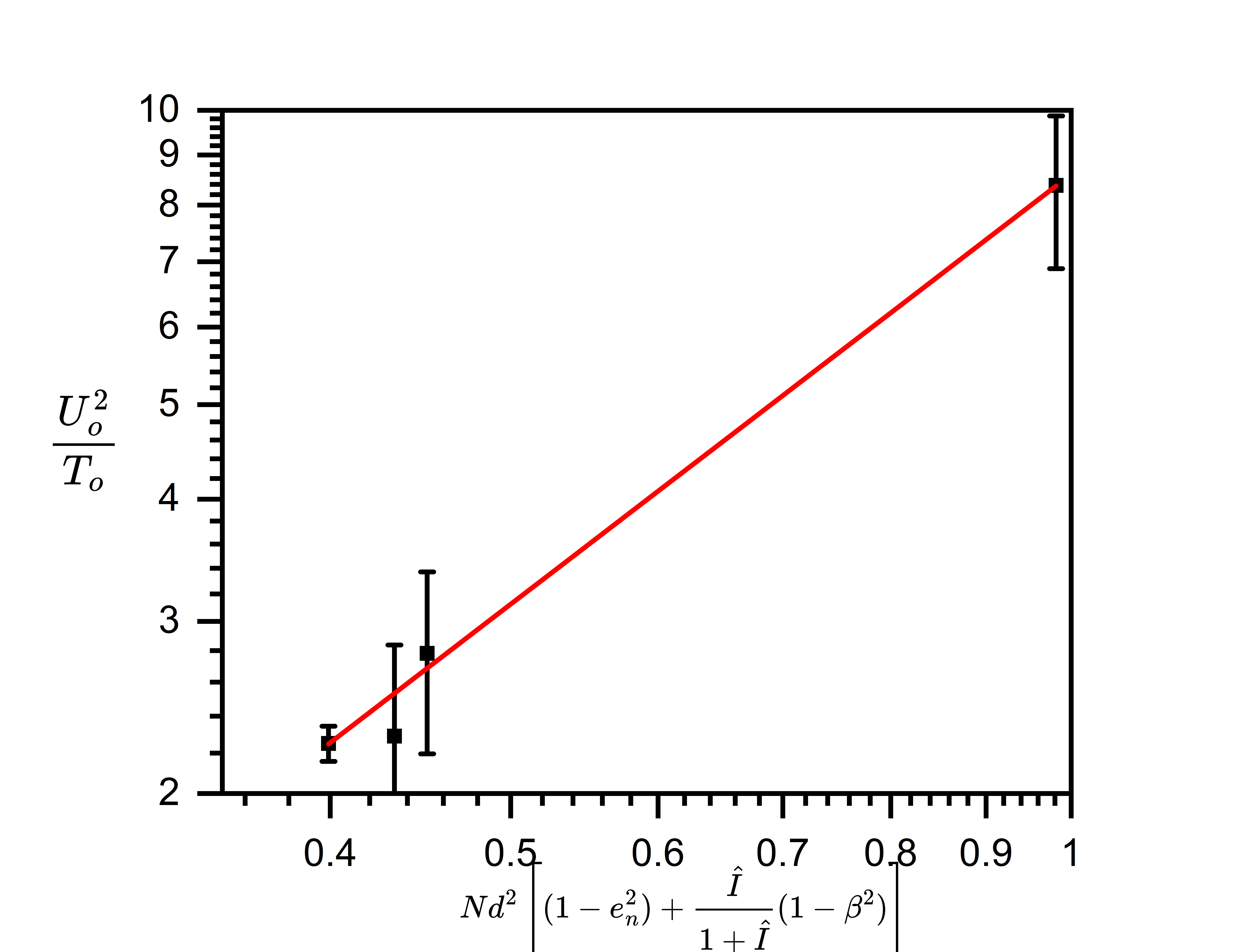}\label{fig:scalenearsmooth}}
     \caption{(a) For perfectly smooth spheres $(\mu = 0)$, (b) For particles with high friction coefficient $(\mu = 0.5, 10$, $\kappa = \frac{2}{7}$, individual contacts were resolved to determine $\beta$ from the post and pre- slip velocities of the contact point, the median value of $\beta$ is used for obtaining the values on the abscissa, (c) for cases with nearly perfectly smooth particles, $\mu <0.1$, $\kappa = \frac{2}{7}, \frac{3}{4}$, $\frac{T_{R,o}}{T_{o}}\ll 1 $, $-1< \beta <0$ }
     \label{fig:Toscale}
 \end{figure*}

Figure \ref{fig:smoothscale} and \ref{fig:roughscale} plot $\frac{U_{o}^{2}}{T_o}$ vs $Nd^2(1-e_{n}^{2})$ for simulations with perfectly smooth particles and $Nd^2\left[(1-e_n^2)+(1-\beta^2)\right]$ for vibro-fluidized particles with high surface roughness, respectively. Results are in excellent agreement with the theory (Eq. ~\ref{eq:Tosmooth}) for $Nd^2(1-e_{n}^{2})<1$ \citep{Kumaran1998jfm}. Eq \ref{eq:Torough} is a direct extension of the leading order theory for rough particles in the dilute vibro-fluidized bed \citep{Kumaran1998jfm, Rao2008}. An excellent agreement between the simulation results and theory for $Nd^2\left[(1-e_n^2)+(1-\beta^2)\right] <1$ is observed. Figure ~\ref{fig:scalenearsmooth} plots $\frac{U_{o}^{2}}{T_o}$ vs. $Nd^2\left[(1-e_n^2)+\frac{\hat{I}}{1+\hat{I}}(1-\beta^2)\right]$ for particles with $\mu <0.1$. Eq. ~\ref{eq:TTrRough} can be rewritten neglecting the last term on the right-hand side, since $\frac{T_{R,o}}{T_o}\ll1$ for these cases. It is worth noting that a simple scaling for $T_{R,o}$ cannot be obtained in this regime. Due to the lack of a unique scaling for $T_o$ for the wide range of roughness, the results are plotted against the dimensional parameter $U_o^2$ in Figure ~\ref{fig:Kfordiffen}.
The height averaged fluctuating kinetic energy, $\mathrm{KE_T}$ and $\mathrm{KE_{R}}$ are determined for $0\le\mu\le10$, three different values of $e_n$, and $\kappa$ and the ratio $k = \frac{\mathrm{KE_T}}{\mathrm{KE_R}}$ is examined in Section \ref{sec:ratio}. The effect of the frictional losses is discussed in Section ~\ref{sec:energy}.

\subsection{Bed height averaged mean fluctuating kinetic energy}
\label{sec:ratio}
\begin{figure*}[htbp]
\centering
\subfloat[a][]{\includegraphics[scale=0.25]{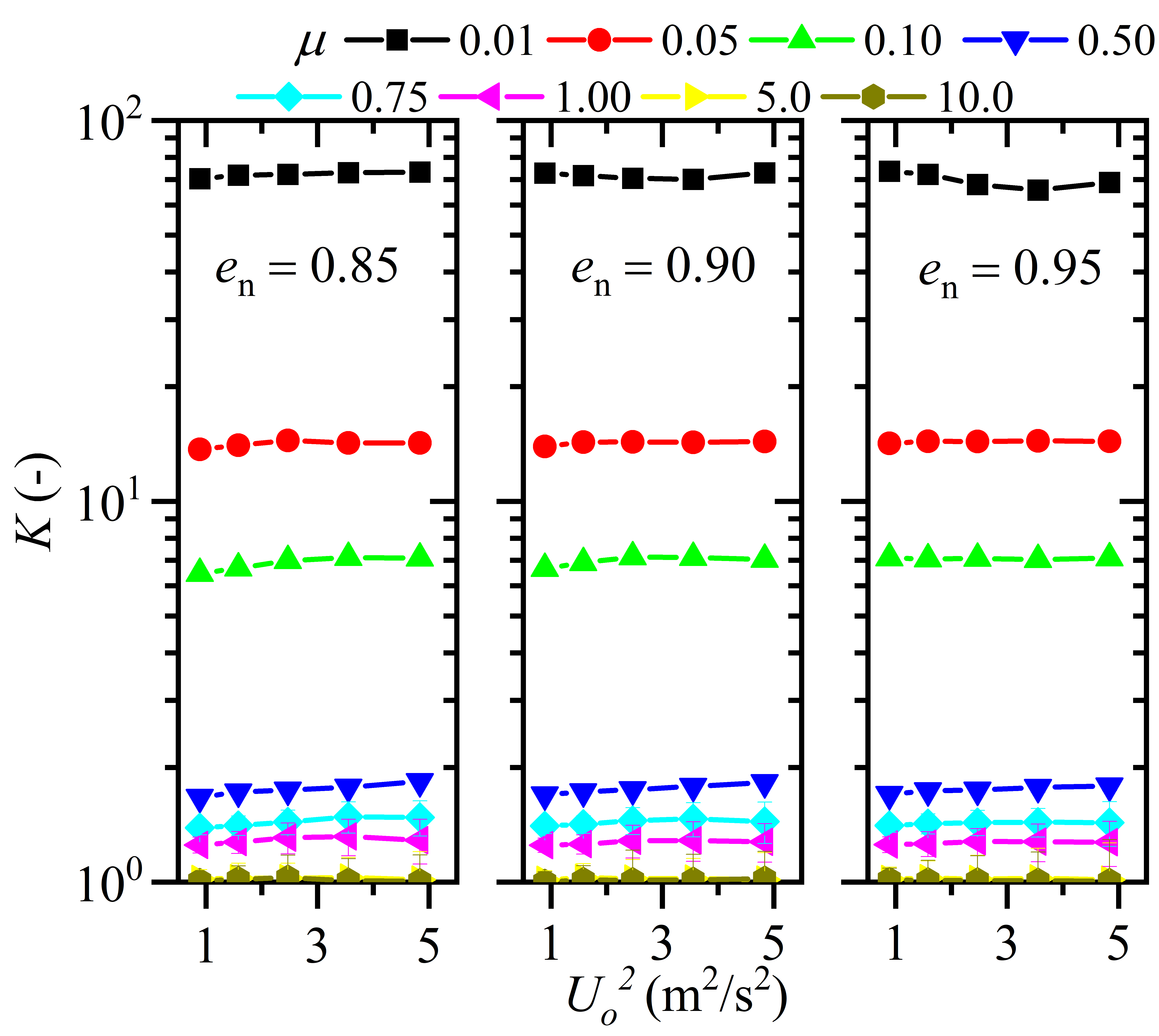}\label{fig:Kfordiffen}}
    \subfloat[b][]{\includegraphics[scale=0.25]{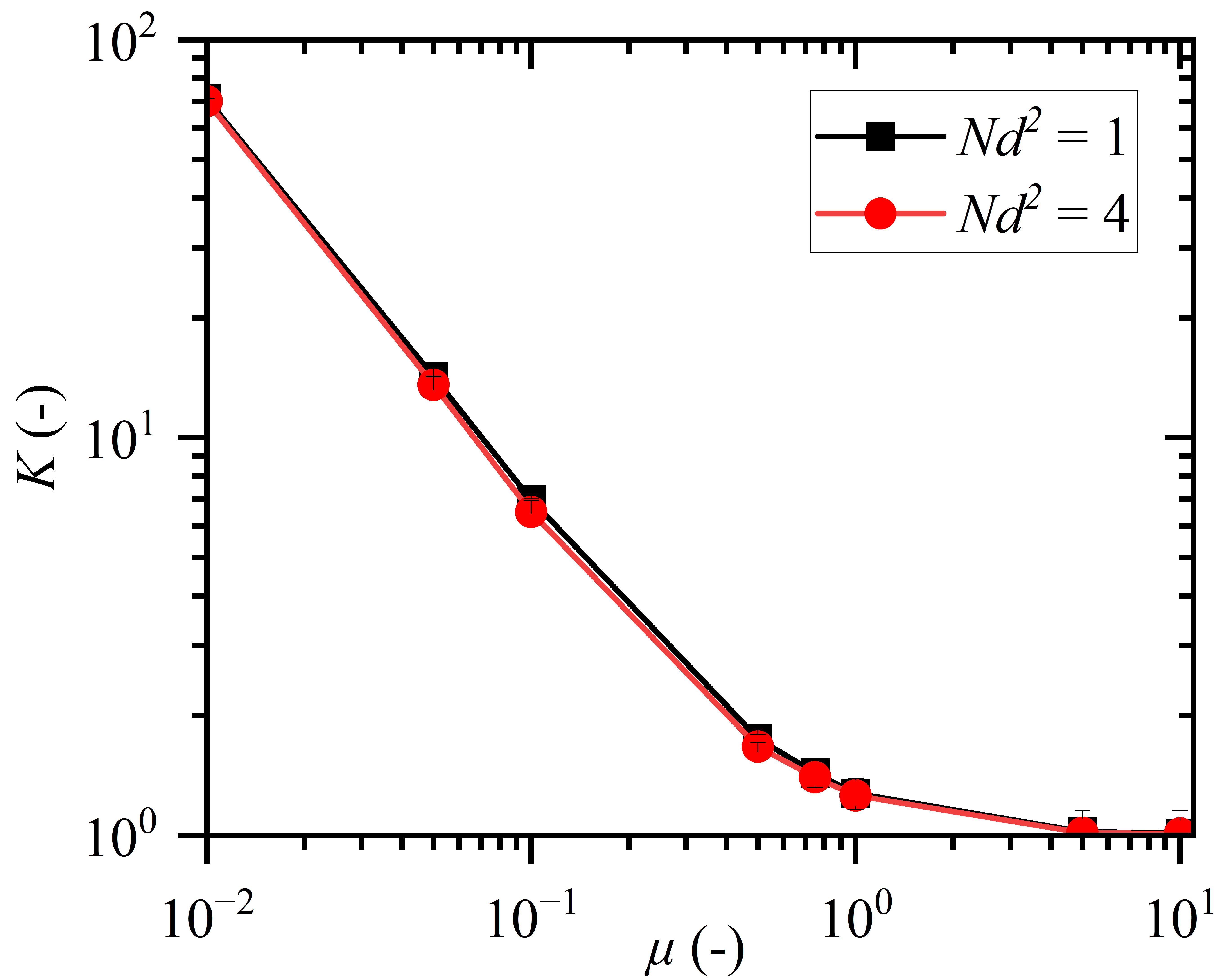}\label{fig:eps_equipartition}}
    
    \subfloat[c][]{\includegraphics[scale=0.25]{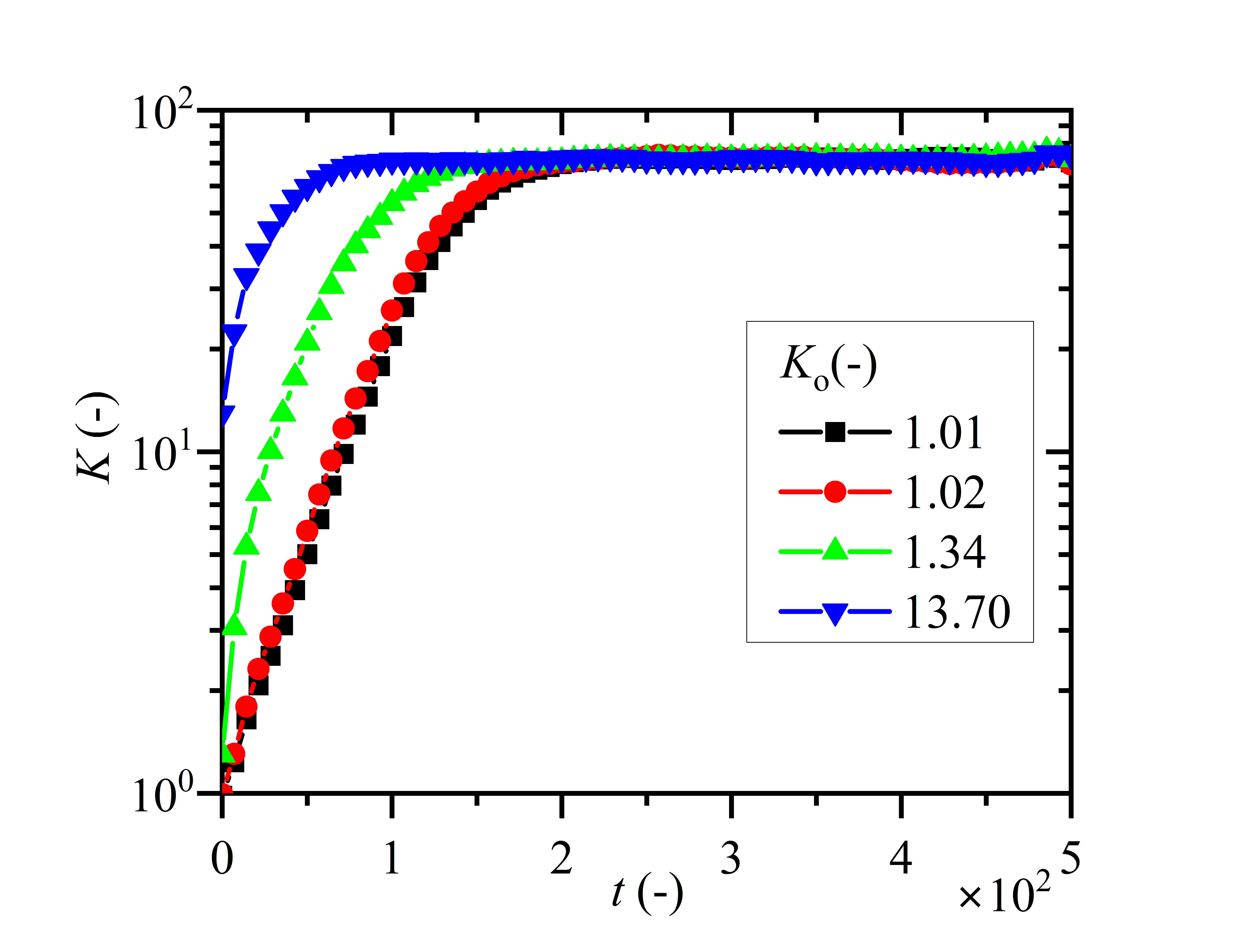}\label{fig:ic_equipartition}}
    \subfloat[d][]{\includegraphics[scale=0.25]{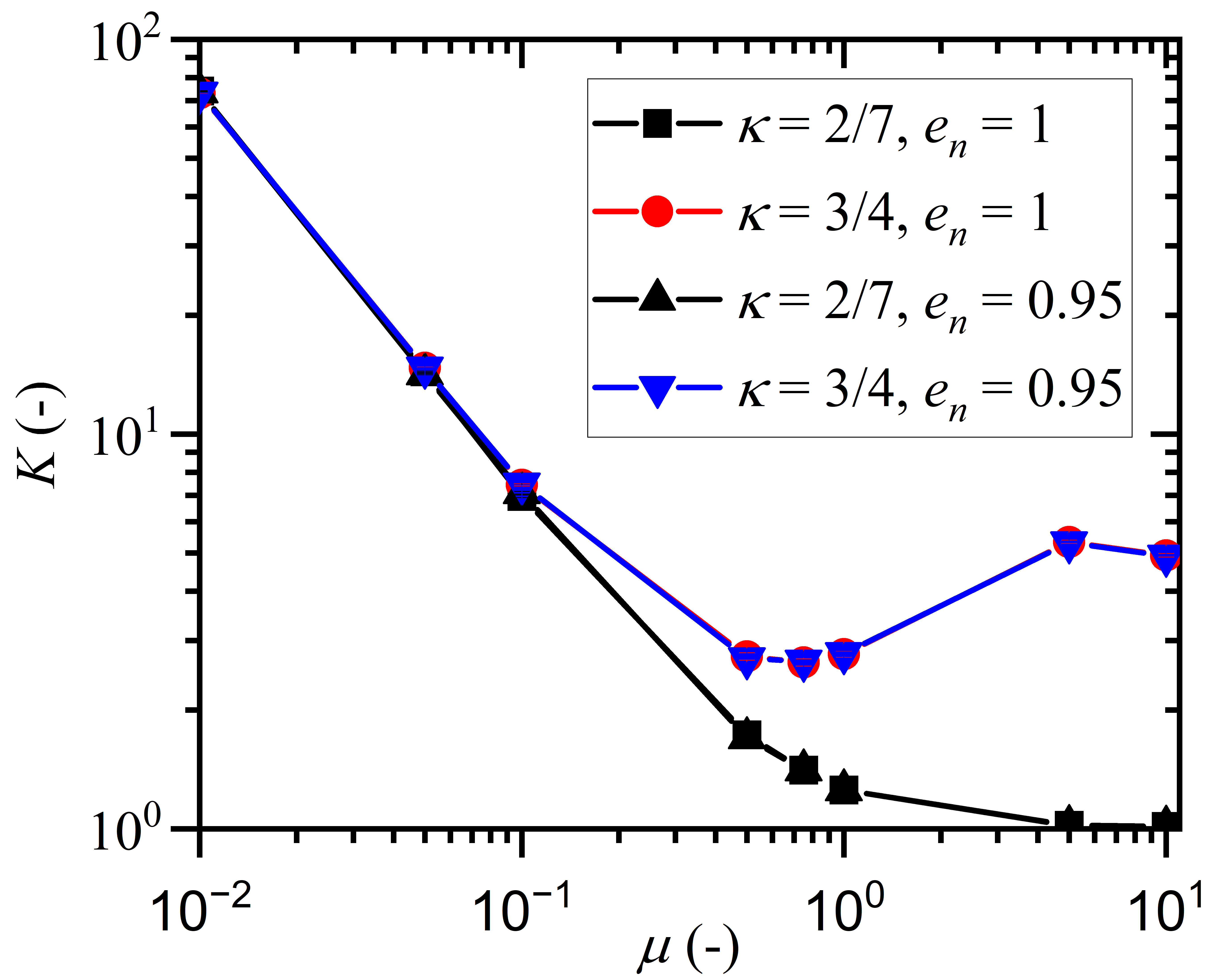}\label{fig:Kforen1}}
\caption{Ratio of translational to rotational fluctuating kinetic energy $(K)$ obtained from the DEM simulation is plotted for different values of $U^2_o$ by varying $\mu$ in the range $0.01$ to $10$ and $e_n = 0.85,0.90$ and $0.95$, keeping $Nd^2 = 4$ constant. (b)The average value of $K$ over $U^2_o$ is plotted against $\mu$ for $(Nd^2)$ = 1, 4 for $e_n = 0.95$. (c) The temporal evolution of $K$ with different initial energy ratios $K_o$ for $\mu=0.01$, $e_n = 0.95$ and $Nd^2 = 4$. (d) Plot of $K$ vs $\mu$ for $\kappa=2/7$ and $3/4$ for $Nd^2=4$ and $e_n =0.95$ and 1.} 
\label{fig:energy_equip}
\end{figure*}

The ratio, $K = \frac{\mathrm{KE_T}}{\mathrm{KE_R}}$ is plotted against $U_o^2$ for $Nd^2 = 4$, $e_n$ = $ 0.85, 0.9, 0.95$ and $\mu$ ranging from 0.01 to 10 in Figure \ref{fig:Kfordiffen}). The plots in the panel suggest that $K$ is independent of base velocity and the normal coefficient of restitution and depends only on the friction coefficient. As the friction coefficient increases, $K$ approaches unity. Figure ~\ref{fig:eps_equipartition} shows the variation
of $K$ with $\mu$ for two different number densities, $Nd^2 = 1, 4$,  and coefficient of restitution $e_n = 0.95$. The ratio of the mean fluctuating rotational to the translational kinetic energy, $K$ is found to be independent of the number density.
The same final $K$ was obtained across simulations with four different initial configurations, each with a distinct value of $K_o = K(t=0)$. This initial condition independence is shown in figure \ref{fig:ic_equipartition}. Figure ~\ref{fig:Kforen1} shows $K$ vs $\mu$ for perfectly smooth particles. In this case, the dissipation is purely due to friction. $K$ decreases monotonically with $\mu$ and plateaus at unity as $\mu$ assumes a very high value, for $\kappa = \frac{2}{7}$. In contrast, for $\kappa=\frac{3}{4}$, the behaviour is non-monotonic; $K$ starts to deviate from the value for $\kappa = \frac{2}{7}$ for $\mu \geq 0.1$. To understand the cause of the deviation, the DEM simulation data are analyzed and presented within the hard-sphere framework. As a first step, the data is processed (a) to identify the contact and determine the effective value of $\beta$ and (b) to determine the terms in Equations ~\ref{eq:delta_re} and \ref{eq:delta_total}.

\subsection{Energy balance during contact}
\label{sec:energy}

\begin{figure*}[htbp]
\centering{
    \subfloat[a][]{\includegraphics[scale=0.25]{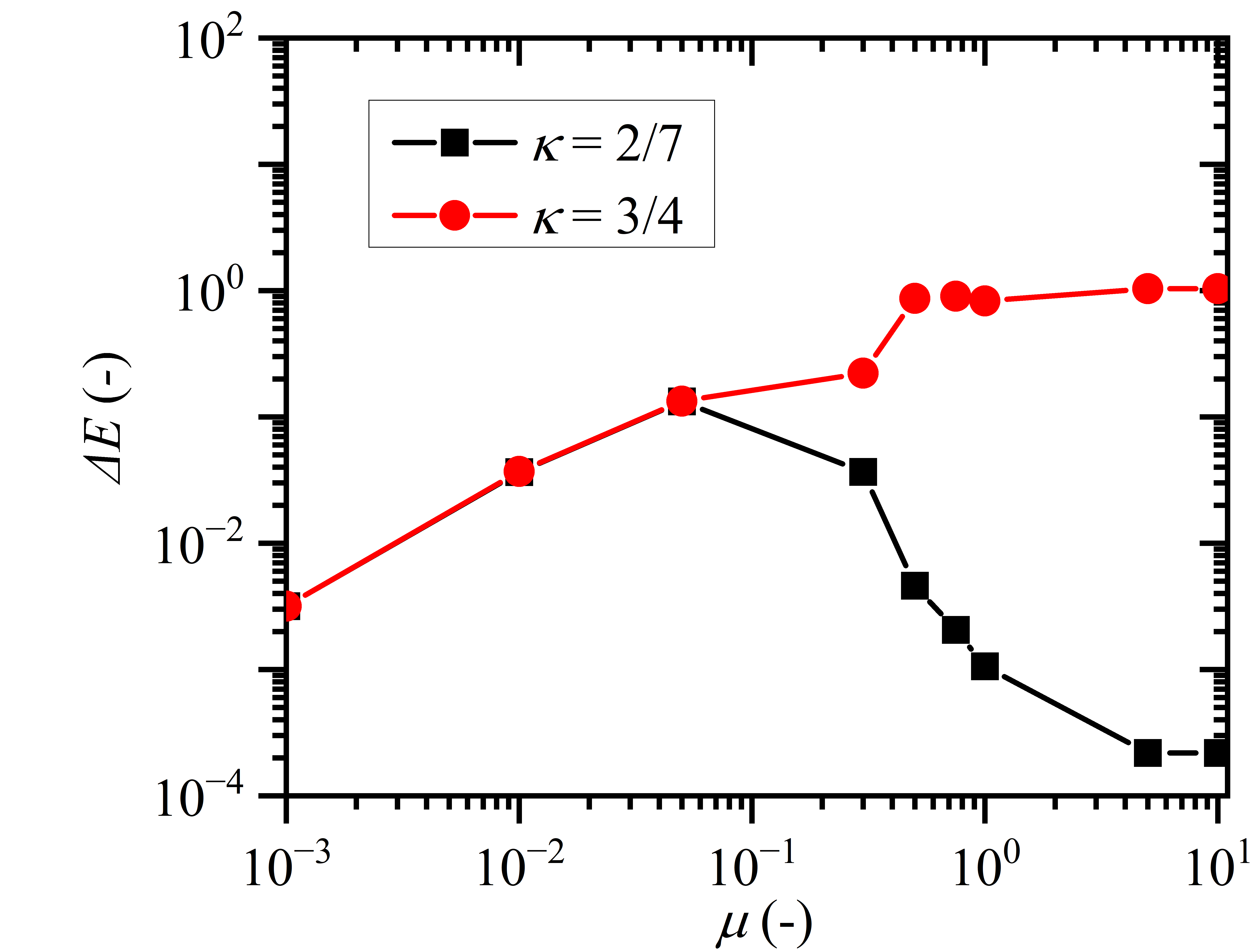}\label{fig:dissipation}}
    \subfloat[b][]{\includegraphics[scale=0.25]{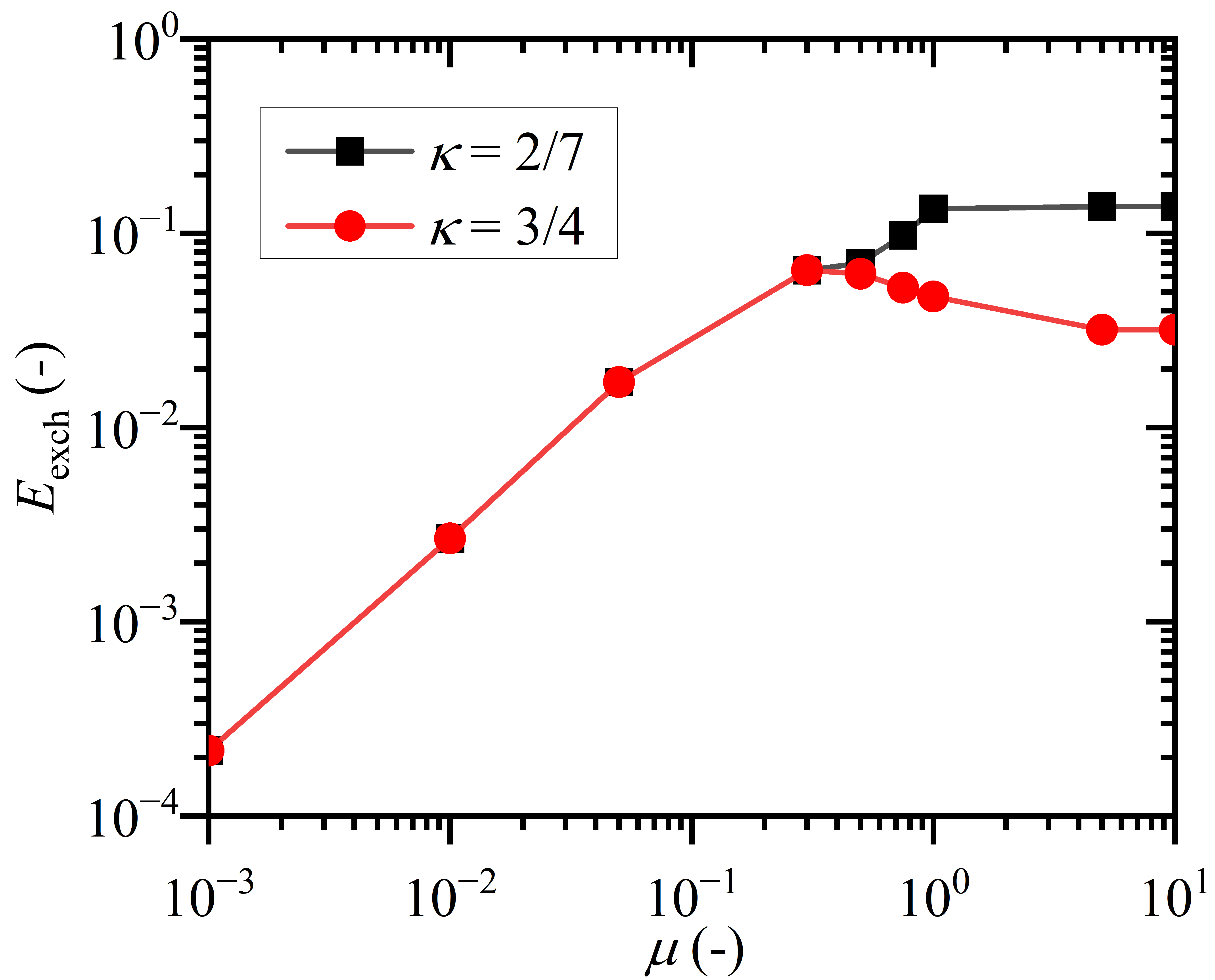}\label{fig:coupling}}\\
    \subfloat[c][]{\includegraphics[scale=0.25]{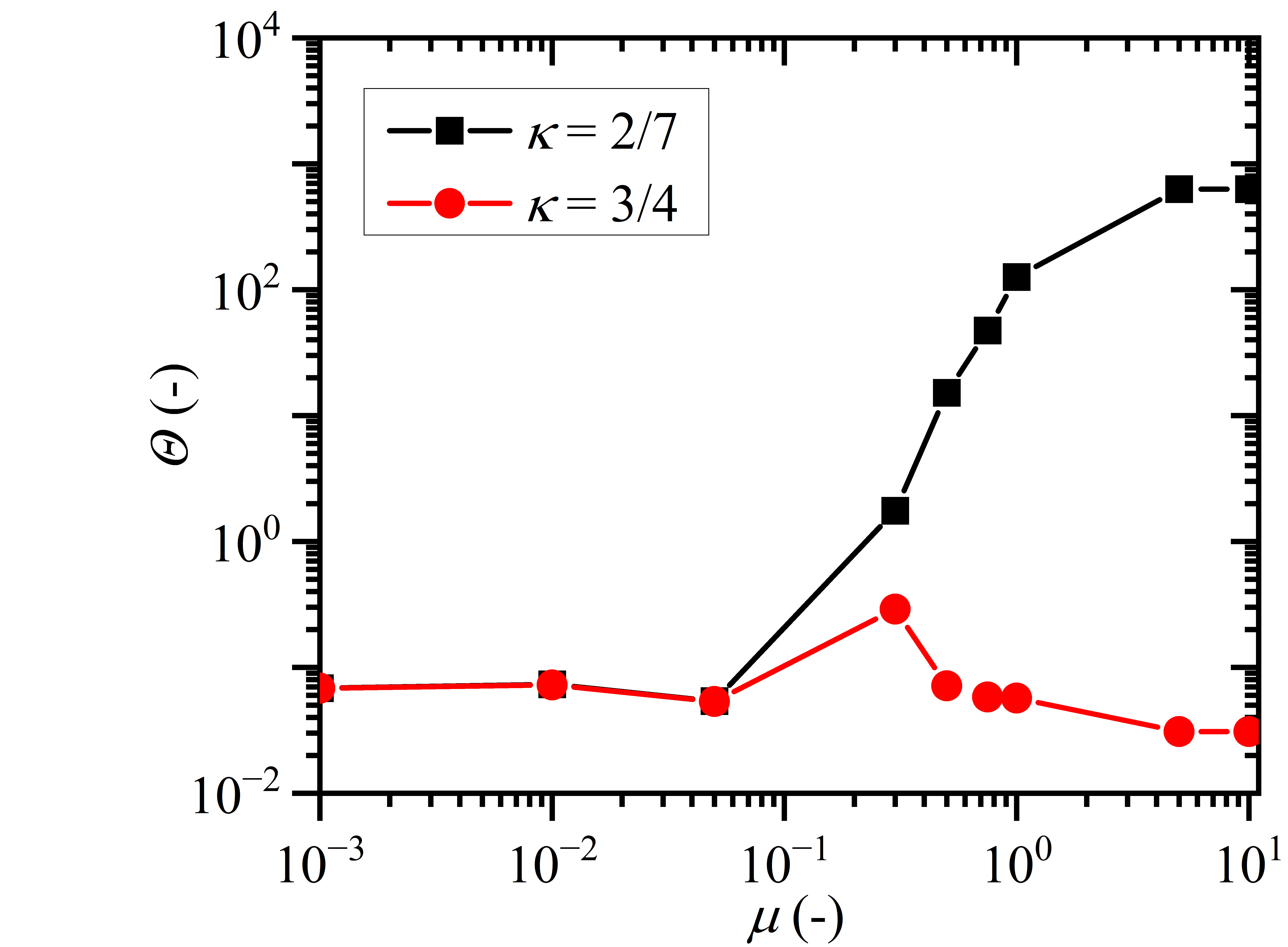}\label{fig:ratio}}}
         
    \caption{(a) $\Delta E = \frac{\hat{I}}{1+\hat{I}}\frac{1-\beta^2}{4}\left ( \hat{r}_{ij} \times \Vec{G}\right ) \ \cdot \left ( \hat{r}_{ij} \times \Vec{G}\right )$ and (b) $ E_{\text{exch}} = \eta_2 \left| \Vec{\omega}_{ij}\cdot \left ( \hat{r}_{ij} \times \vec{G} \right ) \right |$ and (c) $\Theta = \frac{ E_{\text{exch}}}{\Delta E}$ plotted against the friction coefficient $\mu$}\label{fig:energy_terms}. 
    
    \label{fig:force_2_7}
\end{figure*}

The instantaneous positions of the particles are analyzed in a manner similar to that described in \citep{Tiwari2024, SupMat}. Once the contacts are identified and the binary nature of the collisions is verified, the parameter $\beta = -\frac{v_s'}{v_s}$ is determined from the pre- and the post-collision velocities. The energy change $\Delta E$ (Eq. \ref{eq:delta_total}) and $E_{\mathrm{exch}}(=\eta_2 \left| \Vec{\omega}_{ij}\cdot \left ( \hat{r}_{ij} \times \vec{G} \right ) \right |$  are determined for each contact detected.
The median (Q2) of the distribution of $\Delta E$ and $E_{\mathrm{exch}}$ are plotted as a function of $\mu$ in Figure ~\ref{fig:energy_terms}.
Data is collected over $10^3$ configurations from the simulations performed with $\Delta t = \frac{t_c}{100}$.

$\Delta E$ is non-monotonic for $\kappa = \frac{2}{7}$. Dissipation during a collision due to friction increases with $\mu$ up to $\mu = 0.1$, and reduces thereafter. The sliding and sticking regimes are mutually exclusive for $\kappa = \frac{2}{7}$. In the sticking regime, $\beta \approx 1$. In this limit, the collisions are energy conserving (Eq ~\ref{eq:delta_total}. With $\mu$, the fraction of contact in the sticking regime increases, resulting in more energy-conserving contacts. Nearly 75$\%$ of the contact is in the sticking regime for $\mu = 1$ (Figure presented in \cite{SupMat}). In case of $\kappa= \frac{3}{4}$, $\beta < 1$, and the fraction of contact in the stick-slip regime plateaus at 55$\%$ beyond $\mu = 1$. $E_{\mathrm{exch}}$ increases monotonically with $\mu$ before it plateau for $\kappa = \frac{2}{7}$ and reduces for $\kappa = \frac{3}{4}$. The ratio $\Theta = \frac{E_{\mathrm{exch}}}{\Delta E}$ increases sharply with $\mu$ for $\kappa = \frac{2}{7}$, explaining the equipartitioning of fluctuating energy between the translational and the rotational modes at high $\mu$. In contrast, for $\kappa = \frac{3}{4}$, $\Theta < 1$  for all values of $\mu$ and reduces to $\sim 0.03$ for very rough particles ($\mu > 1$). The equipartition is not observed for $\kappa = \frac{3}{4}$ because dissipation exceeds the energy exchange between the translation and rotational modes.

\section{Conclusion}
An assembly of rough, inelastic spherical particles subject to vertical vibration was simulated using the open-source code LAMMPS. The linear spring-dashpot model is used to determine the normal and tangential forces between particles in contact. The normal spring constant is selected such that the collisions are predominantly binary. Two values of $\kappa (=\frac{k_t}{k_n})$ are selected. The time-period of the normal and the tangential contacts are equal for $\kappa =\frac{2}{7}$ and two mutually exclusive regimes of contacts are obtained. The physical interpretation of the stiffness constant as the inverse of the compliance leads to $0.67 \le \kappa < 1$ \citep{Tiwari2024}. The rotational coefficient of restitution obtained from simulations with $0.67 \le \kappa < 1$ is qualitatively similar to the experimentally observed results (\cite{Tiwari2024} and references therein). Results obtained for $\kappa = \frac{3}{4}$ are presented here for detailed discussion. Simulations were also conducted for $\kappa = 0.67$ and $0.95$, and results confirmed that $K = 5.1\pm0.05, 4.91\pm0.04$ respectively. 

The observations from the simulations are:
\begin{enumerate}
\item The equipartition of the mean-squared fluctuating kinetic energy is observed in simulations with $\kappa = \frac{2}{7}$ and for particles with very high friction coefficients. For this range of parameters, $>75\%$ of contacts fall into the energy-conserving sticking regime. 

\item For $\mu \ge 0.3$ and $0.67 \le \kappa < 1$, the ratio of translation to rotational fluctuating kinetic energy $K$ increases beyond 1.5, and the equipartition of energy is not observed. This is because the stick-slip collisions are not energy-conserving.

\end{enumerate}
Non-equipartition of energy between different degrees of freedom is relevant for granular rheology. The results presented here also suggest that the selection of $\kappa$ may be crucial in predicting macroscopic flow behaviour of realistic particles. 

\clearpage

\begin{acknowledgments}
We acknowledge the Indian Institute of Technology Bombay for the licensed version of Grammarly. The software was used to check the manuscript's English grammar. VK was supported by funding from the MHRD and the Science and Engineering Research Board, Government of India (Grant no. SR/S2/JCB-31/2006).
\end{acknowledgments}

\bibliography{GranMat.bib}

@article{McNamara1998,
title = {Energy nonequipartition in systems of inelastic, rough spheres},
author = {McNamara, Sean and Luding, Stefan},
journal = {Phys. Rev. E},
number = {2},
pages = {2247--2250},
volume = {58},
year = {1998}
}

@article{Kosinski2020,
  title = {Extension of the hard-sphere model for particle-flow simulations},
  author = {Kosinski, Pawel and Balakin, Boris V. and Kosinska, Anna},
  journal = {Phys. Rev. E},
  volume = {102},
  issue = {2},
  pages = {022909},
  numpages = {10},
  year = {2020},
  month = {Aug},
  publisher = {American Physical Society},
}

@article{Walton1986,
    author = {Walton, Otis R. and Braun, Robert L.},
    title = "{Viscosity, granular‐temperature, and stress calculations for shearing assemblies of inelastic, frictional disks}",
    journal = {Journal of Rheology},
    volume = {30},
    number = {5},
    pages = {949-980},
    year = {1986},
    month = {10},
    issn = {0148-6055},
}

@unpublished{SupMat,
    author ={A Tiwari and M Bose and V Kumaran} ,
    title = {Supplementary Material},
    year = {2026},
    note = {Supplementary material is available in the ancillary files.}
}

@book{Reif,
    author ={F Reif} ,
    title = {Statistical Physics},
    publisher = {McGraw Hill, New York},
    year ={1965} 
}

@article{ErpenbeckandCohen1988,
    author = {Jerome G Erpenbesk and E G D Cohen },
    title = {Equipartition of energy in a one-dimensional model of diatomic molecules},
    journal = {Phys. Rev. A},
    volume = {38},
    pages = {3054},
    year = {1988}
}

@article{Castillo2020,
author = {Castillo, Gustavo and Merminod, Simon and Falcon, Eric and Berhanu, Michael},
journal = {Phys. Rev. E},
number = {3},
pages = {32903},
pmid = {32289943},
title = {Tuning the distance to equipartition by controlling the collision rate in a driven granular gas experiment},
volume = {101},
year = {2020}
}

@article{Nichol2012,
   author = {Kiri Nichol and Karen E. Daniels},
   issue = {1},
   journal = {Phys. Rev. Lett.},
   pages = {1-5},
   title = {Equipartition of rotational and translational energy in a dense granular gas},
   volume = {108},
   year = {2012},
}

@article{Grasselli2015,
   author = {Y. Grasselli and G. Bossis and R. Morini},
   issue = {2},
   journal = {European Physical Journal E},
   title = {Translational and rotational temperatures of a 2D vibrated granular gas in microgravity},
   volume = {38},
   year = {2015},
}

@article{Potiguar2021,
   author = {Fabricio Q. Potiguar},
   journal = {Physica A: Statistical Mechanics and its Applications},
   pages = {126077},
   title = {On the translational and rotational granular temperatures in periodically excited 2D granular systems},
   volume = {577},
   year = {2021},
}

@article{Wildman2002,
   author = {R. D. Wildman and D. J. Parker},
   issue = {6},
   journal = {Phys. Rev. Lett.},
   pages = {4},
   title = {Coexistence of Two Granular Temperatures in Binary Vibrofluidized Beds},
   volume = {88},
   year = {2002},
}

@article{Paolotti2003,
   author = {D. Paolotti and C. Cattuto and U. Marini Bettolo Marconi and A. Puglisi},
   issue = {2},
   journal = {Granular Matter},
   pages = {75-83},
   title = {Dynamical properties of vibrofluidized granular mixtures},
   volume = {5},
   year = {2003},
}

@article{Eshuis2007,

author = {Eshuis, Peter and van der Weele, Ko and van der Meer, Devaraj and Bos, Robert and Lohse, Detlef},
journal = {Phys. of Fluids},
number = {12},
title = {Phase diagram of vertically shaken granular matter},
volume = {19},
year = {2007}
}

@article{Schafer1996,
author = {Schafer, J. and Dippel, S. and Wolf, D.E.},
journal = {J. Phys. I},
pages = {5--20},
title = {{Force Schemes in Simulations of Granular Materials}},
volume = {6},
year = {1996}
}

@article{Cundall1979,
author = {Cundall, P.A and Strack, O.D.L},
journal = {Geotechnique},
number = {1},
pages = {47--65},
title = {{A discrete numerical model for granular assemblies}},
year = {1979}
}

@article{Reddy2010,
author = {Reddy, K. A. and Kumaran, V.},
journal = {Phys. of Fluids},
number = {11},
title = {{Dense granular flow down an inclined plane: A comparison between the hard particle model and soft particle simulations}},
volume = {22},
year = {2010}
}

@article{Naplekov2023,
   author = {Dmitry M. Naplekov and Vladimir V. Yanovsky},
   issue = {1},
   journal = {Scientific Reports},
   month = {12},
   title = {Distribution of energy in the ideal gas that lacks equipartition},
   volume = {13},
   year = {2023}
}

@article{Puzyrev2024,
   author = {Dmitry Puzyrev and Torsten Trittel and Kirsten Harth and Ralf Stannarius},
   issue = {1},
   journal = {npj Microgravity},
   month = {12},
   publisher = {Nature Research},
   title = {Cooling of a granular gas mixture in microgravity},
   volume = {10},
   year = {2024}
}

@article{Trittel2024,
   author = {Torsten Trittel and Dmitry Puzyrev and Kirsten Harth and Ralf Stannarius},
   issue = {1},
   journal = {npj Microgravity},
   month = {12},
   publisher = {Nature Research},
   title = {Rotational and translational motions in a homogeneously cooling granular gas},
   volume = {10},
   year = {2024}
}

@article{Kumaran1998jfm,
   author = {V. Kumaran},
   journal = {J. of Fluid Mech.},
   pages = {163-185},
   title = {Kinetic theory for a vibro-fluidized bed},
   volume = {364},
   year = {1998}
}

@article{Kumaran1998pre,
   author = {V. Kumaran},
   issue = {5},
   journal = {Phys. Rev. E},
   pages = {5660-5664},
   title = {Temperature of a granular material “fluidized” by external vibrations},
   volume = {57},
   year = {1998}
}

@article{Sunthar1999,
   author = {P. Sunthar and V. Kumaran},
   issue = {2},
   journal = {Phys. Rev. E},
   pages = {1951-1955},
   title = {Temperature scaling in a dense vibrofluidized granular material},
   volume = {60},
   year = {1999}
}

@article{Silbert2001,
   author = {L.E. Silbert and D. Ertaş and G.S. Grest and T.C. Halsey and D. Levine and S.J. Plimpton},
   issue = {5},
   journal = {Phys. Rev. E},
   pages = {14},
   title = {Granular flow down an inclined plane: Bagnold scaling and rheology},
   volume = {64},
   year = {2001}
}

@article{Tiwari2024,
   author = {Alok Tiwari and Sourav Ganguli and Manaswita Bose and V Kumaran},
   month = {12},
   title = {Role of the ratio of tangential to normal stiffness coefficient on the behaviour of vibrofluidised particles},
   journal = {Phys. Rev. E},
   year = {2024}
}

@article{Eastwood2010,
   author = {Michael P Eastwood and Kate A Stafford and Ross A Lippert and Morten {\O} Jensen and Paul Maragakis and Predescu Cristian and Ron O Dror and David E Shaw},
   issue = {7},
   journal = {Journal of Chemical Theory and Computation},
   pages = {2045-2058},
   title = {Equipartition and the Calculation of Temperature in Biomolecular Simulations},
   volume = {6},
   year = {2010}
}

@article{Afek2020,
   author = {Gadi Afek and Alexander Cheplev and Arnaud Courvoisier and Nir Davidson},
   issue = {4},
   journal = {Phys. Rev. A},
   month = {4},
   title = {Deviations from generalized equipartition in confined, laser-cooled atoms},
   volume = {101},
   year = {2020}
}

@book{Rao2008,
   author = {K.K. Rao and P.R. Nott},
   isbn = {9780521571661},
   pages = {490},
   publisher = {Cambridge University Press},
   title = {An Introduction to Granular Flow Hardcover - Version details - Trove},
   year = {2008}
}

@incollection{Walton1993,
  title     = {Particulate Two-Phase Flow},
  author    = {O. R. Walton},
  editor    = {M. Roco},
  chapter   = {25},
  pages     = {884--911},
  publisher = {Butterworth-Heinemann},
  year      = {1992}
}

@article{Maw1976,
author = {N. Maw and J.R. Barber and J.N. Fawcett},
title = {The oblique impact of elastic spheres},
journal = {Wear},
volume = {38},
number = {1},
pages = {101-114},
year = {1976},
issn = {0043-1648},
}

@article{FeitosaandMenon2002,
   author = {Klebert Feitosa and Narayan Menon},
   doi = {10.1103/PhysRevLett.88.198301},
   issn = {198301},
   issue = {19},
   journal = {Phys. Rev. Lett.},
   pages = {4},
   title = {Breakdown of Energy Equipartition in a 2D Binary Vibrated Granular Gas},
   volume = {88},
   year = {2002}
}

\end{document}